\documentclass[epj]{svjourmod}

\usepackage{epsfig}
\usepackage{amsmath,amsfonts,amssymb}
\usepackage{rotating}
\usepackage{citesort}

\newcommand{\rea}[1]{\Re{A_#1}}
\newcommand{\rev}[1]{\Re{V_#1}}
\newcommand{\ubar}{\overline{u}}
\newcommand{\dbar}{\overline{d}}
\newcommand{\sbar}{\overline{s}}
\newcommand{\nubar}{\overline{\nu}}

\newcommand{\lagr}{\mathcal{L}}
\newcommand{\peslash}{{p_\ell\negthickspace\negthickspace\negthickspace/\thickspace}}
\newcommand{\qslash}{{q\negthickspace\negthickspace/\thickspace}}
\newcommand{\order}{\mathcal{O}}
\renewcommand{\Im}[1]{\operatorname{Im}\{#1\}}
\renewcommand{\Re}[1]{\operatorname{Re}\{#1\}}
\renewcommand{\vec}[1]{\mathbf{#1}}
\newcommand{\sN}{\mathcal{N}}
\newcommand{\had}{\operatorname{had}}
\newcommand{\elm}{\operatorname{em}}
\newcommand{\SD}{\operatorname{SD}}
\newcommand{\IB}{\operatorname{IB}}
\newcommand{\degree}{^\circ}
\newcommand{\tee}[1]{\times 10^{#1}}
\newcommand{\xiodd}{\vec{q}\cdot(\vec{p}_\ell\times \vec{p'})/M_K^3}
\newcommand{\even}{{\operatorname{even}}}
\newcommand{\odd}{{\operatorname{odd}}}
\newcommand{\oneloop}{\begin{rotate}{90}$\;$(2-$\pi$-cut)\end{rotate}}
\newcommand{\twoloop}{\begin{rotate}{90}$\;$(3-$\pi$-cut)\end{rotate}}
\newcommand{\av}[1]{\langle {#1}\rangle}
\newcommand{\mc}[1]{\multicolumn{#1}{l}}
\newcommand{\Eg}{E_\gamma^*}
\newcommand{\Ecut}{E_\gamma^{\rm cut}}
\newcommand{\Ee}{{E_e^*}}
\newcommand{\te}{\theta_{e\gamma}^*}
\newcommand{\tecut}{\theta_{e\gamma}^{\rm cut}}

\def\kl3g{K_{\ell3\gamma}}
\def\beq{\begin{equation}}
\def\eeq{\end{equation}}
\def\bea{\begin{eqnarray}}
\def\eea{\end{eqnarray}}

\begin{document}


\title{$\boldsymbol{T}$-odd correlations in radiative $\boldsymbol{K^+_{\ell3}}$ decays\\ 
  and chiral perturbation theory}
\titlerunning{$T$-odd correlations in radiative $K^+_{\ell3}$ decays and chiral perturbation theory}

\author{
Eike H. M\"uller\inst{1} 
\thanks{Electronic address:  \texttt{emueller@itkp.uni-bonn.de}}
\and
Bastian Kubis\inst{1} 
\thanks{Electronic address:  \texttt{kubis@itkp.uni-bonn.de}}
\and 
Ulf-G. Mei{\ss}ner\inst{1,2}
\thanks{Electronic address:  \texttt{meissner@itkp.uni-bonn.de}}
}                     

\institute{
Helmholtz-Institut f\"ur Strahlen- und Kernphysik, Universit\"at Bonn, 
Nu{\ss}allee 14--16, D-53115 Bonn, Germany
\and
Forschungszentrum J\"ulich, Institut f\"ur Kernphysik (Theorie), 
D-52425  J\"ulich, Germany
}

\date{}

\abstract{
The charged kaon decay channel $K^+_{\ell 3\gamma}$ allows for 
studies of direct $CP$ violation,
possibly due to non-standard mechanisms, 
with the help of $T$-odd correlation variables.
In order to be able to extract a $CP$-violating signal from experiment, 
it is necessary to understand all possible standard model phases 
that also produce $T$-odd asymmetries. 
We complement earlier studies by considering strong interaction phases
in hadronic structure functions that appear at higher orders in chiral perturbation theory,
and we compare our findings to other potential sources of asymmetries.
\PACS{
      {13.20.Eb}{Radiative semileptonic decays of $K$ mesons}
      \and
      {11.30.Er}{Charge conjugation, parity, time reversal, and other discrete symmetries}
      \and
      {12.39.Fe}{Chiral Lagrangians}
     }
}

\maketitle

\section{Introduction}\label{intro}

The decays of charged kaons may serve as an excellent system to study direct $CP$ violation in particle physics.
As this is expected to be totally negligible in particular for semileptonic kaon decays
within the standard model, the investigation of such channels  may even give access to 
non-standard $CP$-violating mechanisms, or limits thereon.

\begin{sloppypar}
As the simplest $CP$-violating observable, namely the charge asymmetry of the decay widths
$\Gamma(K^+\to f)-\Gamma(K^-\to \bar f)$, can only be non-vanishing in the presence
of at least two weak amplitudes with different re-scattering phases~\cite{D'Ambrosio96}, 
it has been suggested to resort to $T$-odd correlations in order to test $CP$
in the context of semileptonic charged kaon decays.  
In the decays $K^+\to \pi^0\mu^+\nu_\mu$ ($K_{\mu 3}$) or  
$K^+\to \mu^+\nu_\mu \gamma$ ($K_{\mu 2\gamma}$)
(see e.g.\ Refs.~\cite{Zhi80,Bel91,Chen97,Efr00,Diw01,Ani03,Ani04,Bra05,Abe06} and references therein)
the transverse muon polarization is a suitable observable that is odd under time reversal.
\end{sloppypar}

In experiments that do not have access to the lepton polarization, however, 
a $T$-odd correlation can still be constructed as a triple scalar product $\xi$ of three independent momenta
in decay channels with at least four particles in the final state, 
such as $K^+\to \pi^+ \pi^- \ell^+\nu_\ell$  ($K_{\ell 4}$)~\cite{Retico02} 
or $K^+\to \pi^0 \ell^+\nu_\ell \gamma$ ($\kl3g$), 
which we (re-)investigate in this article.
Violation of $T$-invariance in such channels will manifest itself in $\xi$-odd contributions to the 
differential width $d\Gamma/d\xi$, or in the asymmetry $N(\xi>0)-N(\xi<0)$.
Experiments studying this effect in $\kl3g$ have started producing results recently~\cite{Obr01,Tch05,Bol05} 
but still lack the necessary precision to test theoretical predictions. 

An investigation of the $T$-odd asymmetries in $\kl3g$ that arise in extensions of the standard model 
has first been performed in Ref.~\cite{Bra03}. 
In principle, if both the decay of the $K^+$ and its charge conjugate mode for the $K^-$ 
are measured and their asymmetries appropriately combined, these effects 
due to complex non-standard couplings can be unambiguously extracted.
However, if only the $K^+$ \emph{or} the $K^-$ channel are considered, 
there are also contributions to the asymmetry from within the standard model 
due to final state interaction phases, 
which have to be accounted for when determining the part attributed to new physics.
The effects of photon loops that emerge as rescattering effects of the photon
and the charged lepton have been investigated in Ref.~\cite{Bra02}.
Another source, however, which so far seems to have been completely ignored, are phases due
to imaginary parts in the (hadronic) structure functions in $\kl3g$.
These structure functions can be calculated systematically 
in the framework of chiral perturbation theory (ChPT), and they
were shown to be purely real at leading (non-trivial) order ($\order(p^4)$)~\cite{Bij93}, 
but as it was pointed out in Ref.~\cite{Gas05} 
(for the corresponding neutral kaon decay mode), imaginary parts arise
beyond leading order, starting at $\order(p^6)$. 
There is no {\it a priori} reason why these should be neglected compared to electromagnetic phases;
in fact, although of relatively high order in the chiral expansion, they lack the
suppression by $\alpha \sim 1/137$.

In this article, we will clarify the situation by a thorough investigation of 
the imaginary parts of the structure functions in ChPT.
We lay out some necessary formalism in Sect.~\ref{sec:phenomenology}.
We present our main results, the analytic forms of the leading cut contributions 
as well as numerical results for the $T$-odd asymmetries, in Sect.~\ref{sec:ChPT}. 
In Sect.~\ref{sec:comparison} we compare our findings to the
asymmetries due to photon loops and make some comments on the size and structure
of beyond-the-standard-model contributions.  
Section~\ref{sec:summ} contains our conclusions.
Some more technical aspects are relegated to the appendices.

\section{Phenomenology of $\boldsymbol{K^+_{\ell3\gamma}}$ decays}\label{sec:phenomenology}

We briefly review the phenomenology of the decay
\beq
  K^+(p) \,\to\, \pi^0(p')\,\ell^+(p_\ell)\,\nu_\ell(p_\nu)\,\gamma(q)  \quad \bigl[\kl3g^+\bigr]\,,~
  \ell = e,\,\mu\,, 
\eeq
which has been done in more detail for the corresponding decay of neutral kaons 
in Ref.~\cite{Gas05}.

\subsection{Matrix element\label{sec:matrixelement}}

The $\kl3g^+$ matrix element can be written as
\begin{eqnarray}
  T(K^+_{\ell3\gamma}) &=& \frac{G_F}{\sqrt{2}}eV_{us}^*\epsilon^\mu(q)^* \notag\\
  &\times&\biggl[
  (V_{\mu\nu}-A_{\mu\nu})\nubar(p_\nu)\gamma^\nu (1-\gamma_5)\ell(p_\ell) \label{eqn:matrixelement}
  \\
  &+&\frac{F_\nu}{2p_\ell q}\nubar(p_\nu)\gamma^\nu (1-\gamma_5)
  (m_\ell-\peslash-\qslash)\gamma_\mu\ell(p_\ell)
  \biggr] ~,\notag 
\end{eqnarray}
where the hadronic tensors $V_{\mu\nu}$ and $A_{\mu\nu}$ are given by
\begin{eqnarray}
I_{\mu\nu} &=& i\int d^4x\; e^{iqx}\langle\pi^0(p')|
    T\{V_\mu^{\elm}(x)I^{\had}_\nu(0)\}|K^+(p)\rangle ~,\notag\\
I &=& V,\,A ~,\label{eqn:formfactors}
\end{eqnarray}
with
\begin{eqnarray}
V_\nu^{\had} &=& \sbar \gamma_\nu u ~,\quad A_\nu^{\had} = \sbar \gamma_\nu \gamma_5 u ~,\notag\\
V_\mu^{\elm} &=& (2\ubar\gamma_\mu u-\dbar\gamma_\mu d-\sbar\gamma_\mu s)/3 ~,
\end{eqnarray}
whereas $F_\nu$ is the  $K^+_{\ell3}$ vector correlator
\beq
  F_\nu = \langle\pi^0(p')|V_\nu^{\had}|K^+(p)\rangle ~.
\eeq
The tensors $V_{\mu\nu}$ and $A_{\mu\nu}$ satisfy the electromagnetic Ward identities
\begin{xalignat}{2}
  q^\mu V_{\mu\nu} &= F_\nu ~,&\quad  q^\mu A_{\mu\nu} &= 0 ~,\label{eqn:emWard}
\end{xalignat}
which guarantee gauge invariance of the amplitude \eqref{eqn:matrixelement}.

The $\kl3g^+$ amplitude can be decomposed into (separately gauge invariant)
inner bremsstrahlung (IB) and structure dependent (SD) parts.
According to Low's theorem~\cite{Low58}, the IB terms, which comprise
the part of the amplitude non-vanishing for small photon momenta
(and in particular the infrared divergent pieces),
is given entirely in 
terms of the $K^+_{\ell3}$ form factors $f_+$, $f_1$ defined by
\begin{eqnarray}
  F_\nu(t) &=& \frac{1}{\sqrt{2}}\Big(2p'_\nu f_+(t)+(p-p')_\nu f_1(t)\Big)
\end{eqnarray}
with $t=(p-p')^2$.
This splitting of the matrix element implies a corresponding splitting of the 
hadronic tensors $V_{\mu\nu}$ and $A_{\mu\nu}$.
The decomposition of the vector correlator reads
\begin{eqnarray}
  V_{\mu\nu} &=& V_{\mu\nu}^{\IB} + V_{\mu\nu}^{\SD} ~,
\end{eqnarray}
where the IB piece is chosen such that
\begin{xalignat}{2}
  q^\mu V_{\mu\nu}^{\IB} &= F_\nu(t)~, &  q^\mu V_{\mu\nu}^{\SD} &= 0 ~.
\end{xalignat}
$V_{\mu\nu}^{\IB}$ can be shown~\cite{GKMSxx} 
(see also Ref.~\cite{Gas05} for a corresponding derivation for $\kl3g^0$,
and Refs.~\cite{FFS70a,FFS70b} for slightly different representations)
to be given in terms of $f_+$ and $f_1$ as follows:
\begin{eqnarray}
  V_{\mu\nu}^{\IB} &=& \frac{1}{\sqrt{2}}
  \bigg[
  \frac{p_\mu}{pq}\bigl(2p'_\nu f_+(W^2) + W_\nu f_1(W^2)\bigr)\notag\\ &&
  +\;\;\frac{W_\mu}{qW}\bigl(2p'_\nu \Delta f_++W_\nu \Delta f_1\bigr)
  +g_{\mu\nu}f_1(t)
  \bigg]\notag ~,\\
  \Delta f_i &=& f_i(t)-f_i(W^2) ~,\quad i = +,1 ~,
\end{eqnarray}
where $W = p-p'-q$. 
The structure dependent part can be expressed 
in terms of four scalar functions $V_i$, $i=1\ldots4$, 
in a basis of gauge invariant tensors according to
\begin{eqnarray}
  V_{\mu\nu}^{\SD} &=& \frac{1}{\sqrt{2}}\Bigl[
  V_1 \,(p'_\mu q_\nu-p'q\,g_{\mu\nu}) +
  V_2 \,( W_\mu q_\nu- qW\,g_{\mu\nu}) \notag\\
  &&\qquad + V_3 \,\bigl(qW \, p'_\mu W_\nu  - p'q \, W_\mu W_\nu \bigr) \notag\\
  &&\qquad + V_4 \,\bigl(qW \, p'_\mu p'_\nu - p'q \, W_\mu p'_\nu\bigr)
  \Bigr] ~.
\end{eqnarray} 
The axial correlator $A_{\mu\nu}$ consists of structure dependent parts only,
and can also be expressed in terms of four scalar functions $A_i$, $i=1\ldots4$,
\begin{eqnarray}
  A_{\mu\nu} 
  &=& \frac{i}{\sqrt{2}}\biggl[
  \epsilon_{\mu\nu\rho\sigma}\bigl(A_1 \,p'^\rho q^\sigma + A_2 \,q^\rho W^\sigma\bigr) \label{eqn:Aidefinition}\\
  && +\, \epsilon_{\mu\lambda\rho\sigma} \, p'^\lambda q^\rho W^\sigma
  \biggl( \frac{A_3}{M_K^2-W^2}\,W_\nu+A_4\,p'_\nu\biggr)
  \biggr] ~.\notag
\end{eqnarray}
(We use the convention $\epsilon_{0123} = +1$.)
Note that in comparison to Refs.~\cite{Bij93,Gas05}, we have, for convenience, factored
out the kaon pole explicitly in the definition of the structure function $A_3$. 

\subsection{Kinematics}\label{sec:kinematics}

The functions $V_i$ and $A_i$ depend on three independent
Mandelstam variables $s$, $t$, and $u$ defined by
\begin{xalignat}{3}
  s &= (p'+q)^2 ~, &
  t &= (p-p')^2 ~, &
  u &= (p-q)^2 ~,
\end{xalignat}
where $s+t+u=M_K^2+M_\pi^2+W^2$.
The precise kinematic limits for these are given in Ref.~\cite{Gas05}.
In order to describe the full kinematics of the $\kl3g^+$ decay,
two more variables are needed, e.g.
\beq
p p_e/M_K = \Ee ~,\quad x = p_e q/M_K^2 ~.
\eeq
One often works in the rest frame of the decaying 
kaon.\footnote{We label quantities in this frame by an upper index `$*$'.}
In order to tame the infrared singularity for small photon energies $\Eg$, 
it is necessary to cut on this variable when calculating observables,
\beq
\Eg \ge \Ecut ~.
\eeq
In addition, there is a near-singularity in the electron channel when $x$ becomes
small for collinear photon and positron momenta, 
\beq
 p_eq =  x \,M_K^2 = \Eg \left(\Ee-\sqrt{\Ee^2-m_e^2}\cos\te \right) ~,
\eeq
therefore in this channel one also cuts on the angle
\beq
  \te \ge \tecut~.
\eeq
We use the ``standard cuts'' $\Ecut = 30$~MeV, 
$\tecut = 20\degree$ most of the time in this work,
except for Sect.~\ref{sec:cutdep} where we explicitly study
the dependence of the $T$-odd asymmetry on these cuts.

\subsection{The asymmetry $\boldsymbol{A_\xi}$}

\begin{sloppypar}
The spin summed squared matrix element can be expressed in terms of the 
functions $V_i$, $A_i$, and the $K_{\ell3}$ form factors $f_+$, $f_1$:
\begin{eqnarray}
  \sN^{-1}\sum_{\operatorname{spins}} |T|^2 &=& 
  \sum_{f,f'} a_{ff'}\;ff' + \sum_{f,I} b_{fI}\;f\Re{I} \label{eqn:Tsquared}\\
  &+& \xi\cdot\sum_{f,I} c_{fI}\;f\Im{I}
+\order(V_i^2,A_i^2,A_iV_j) ~, \notag 
\end{eqnarray}
where
\bea
  f,f' &\in& \{f_+,\, \delta f_+,\, f_1,\, \delta f_1\} ~, \quad
  \delta f_{+,1} \equiv \frac{M_K^2}{qW}\, \Delta f_{+,1} ~, \notag\\
  I &\in& \{V_i, A_i \} ~, \quad i=1\ldots4~,
\eea
and the $T$-odd variable $\xi$ is given by\footnote{The sign of the definition 
\eqref{eqn:xidef} is opposite to Appendix~B in Ref.~\cite{Gas05}, but
agrees with the convention used in Refs.~\cite{Bra02,Bra03}.  We are grateful 
to V.~V.~Braguta for confirming this point.}
\bea
\xi &=& \epsilon_{\mu\nu\alpha\beta}\,p^\mu q^\nu p_\ell^\alpha p'^\beta / M_K^4 \,\notag\\
&\stackrel{{\rm CMS}}{=}& \xiodd   ~,\label{eqn:xidef}
\eea
where the last form refers to the center-of-mass (CMS) system for the decay,
the kaon rest frame.
The functions $a_{ff'}$, $b_{fI}$, $c_{fI}$ are kinematical factors that depend on 
scalar products of the final particle momenta. 
We spell out the factors relevant for the $\xi$-odd part of the squared matrix element
in Appendix~\ref{app:factorsSM}.
Finally the normalization factor in~\eqref{eqn:Tsquared} is
\begin{eqnarray}
  \sN = 8\pi\alpha G_F^2|V_{us}|^2 M_K^2 ~.
\end{eqnarray}
\end{sloppypar}

The first part of the squared matrix element \eqref{eqn:Tsquared},
which contains neither $V_i$ nor $A_i$, only consists of
inner bremsstrahlung and is the dominant part e.g.\ for the partial decay width. 
The other terms  arise because of interference of the IB and SD pieces. 
It was found in Ref.~\cite{Gas05} that these contributions are
suppressed by two orders of magnitude compared to the ${\IB}^2$ part in $K_{e3\gamma}^0$, and we find
the same suppression in the decay described in this paper. Contributions of 
$\order(V_i^2, A_i^2, V_iA_j)$ are suppressed even further, which is why we neglect them.
We remark that the kinematical factors that multiply $f_1$, $\delta f_1$, $V_3$, and
$A_3$ are suppressed by $(m_\ell/M_K)^2$; hence these terms can be dropped in
the electron channel.

It is crucial to note that the imaginary parts of the functions $V_i$, $A_i$ are always multiplied by
$\xi$ in \eqref{eqn:Tsquared}.

We can compute the differential width $d\Gamma/d\xi$ with respect to $\xi$.
As $\xi^2$ can be expressed in terms of scalar products of the final particle momenta, 
we can split the differential width into a $T$-even and a $T$-odd part,
\begin{eqnarray}
  \frac{d\Gamma}{d\xi} &=& \rho_\even + \xi\cdot \rho_\odd ~,
\end{eqnarray}
where $\rho_\even$ and $\rho_\odd$ only depend on scalar products of final particle momenta. 
To compare the relative size of both contributions it is convenient to define the asymmetry
\beq
  A_\xi = \frac{N_+-N_-}{N_++N_-} \quad \text{with} \quad
 N_\pm=\int_{\xi \gtrless 0} d\Gamma ~. \label{eqn:defAxi}
\eeq

\section{Imaginary parts and asymmetry in ChPT}\label{sec:ChPT}

\subsection{ChPT up to $\boldsymbol{\order(p^4)}$}\label{sec:ChPTOp4}

The structure functions $V_i$ and $A_i$ have been evaluated in the framework of 
chiral perturbation theory in Ref.~\cite{Bij93} up to $\order(p^4)$. 
For the axial amplitude the only contributions come from tree level diagrams with 
vertices from the Wess--Zumino--Witten anomalous Lagrangian~\cite{WZ71,Wi83}
\beq
  A_1 = -4 A_2 = A_3 = -\frac{1}{2\pi^2 F^2}~, \quad
  A_4 = 0 ~
\eeq
(where $F$ is the [pseudoscalar] meson decay constant),
whereas the $V_i$ are obtained by evaluating $\lagr^{(4)}$ counterterms and loops with $\lagr^{(2)}$ vertices. 
However, all cuts in the loop diagrams lie well outside the physical region. 
\begin{figure}
  \centering
  \epsfig{width=0.6\linewidth,file=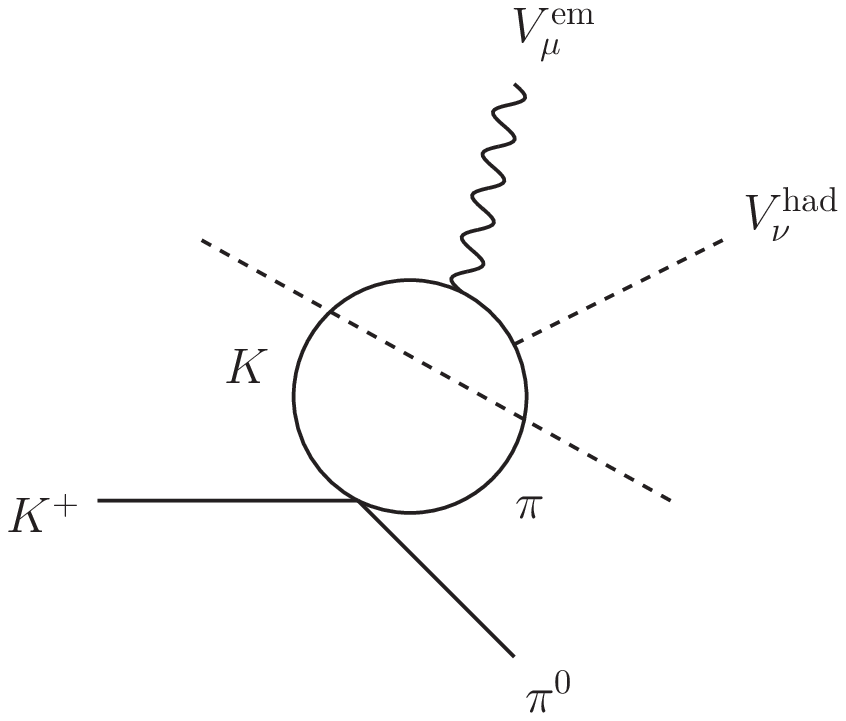}
  \caption{Example of a $\order(p^4)$ loop diagram with cuts for \mbox{$t\ge(M_K+M_\pi)^2$}
           and \mbox{$W^2\ge(M_K+M_\pi)^2$}.}
\label{fig:nlocut}
\end{figure}
To see this consider, for example, the $t$-channel diagram in Fig.~\ref{fig:nlocut}, 
which will develop imaginary parts for $t\ge(M_K+M_\pi)^2$ or $W^2\ge(M_K+M_\pi)^2$. 
The same is true for all $\order(p^4)$ diagrams, which only have cuts in the variables $t$ and $W^2$, 
as, because of strangeness conservation, one always has to cut (at least) one kaon line. 
This implies that the structure functions are real in the physical region.
As shown in Ref.~\cite{GKMSxx} (and similarly for the neutral $\kl3g$ decay in Ref.~\cite{Gas05}),
the structure functions altogether vary only very little over phase space at $\order(p^4)$ and therefore,
to a very good approximation, can be replaced by their average values $\av{V_i}$ as
given in Table~\ref{tab:Vcentralvalues}.
\begin{table}
\caption{Average values of $V_i$, $A_i$ at $\order(p^4)$ (in units of $M_K$).}
\label{tab:Vcentralvalues}
\centering
\renewcommand{\arraystretch}{1.2}
\begin{tabular}{lllll}
\hline\noalign{\smallskip}
$\av{V_1}$ & $-1.24$        && $\av{A_1}$ & $-1.19$\\
$\av{V_2}$ & $-0.19$        && $\av{A_2}$ & $\phantom{-}0.30$\\
$\av{V_3}$ & $-0.02$        && $\av{A_3}$ & $-1.19$\\
$\av{V_4}$ & $\phantom{-}0$ && $\av{A_4}$ & $\phantom{-}0$\\
\noalign{\smallskip}\hline
\end{tabular}
\renewcommand{\arraystretch}{1.0}
\end{table}
The reality of the structure functions implies that the squared matrix element \eqref{eqn:Tsquared} 
does not contain terms that are proportional to $\xi$, and hence the asymmetry $A_\xi$ will be zero 
at this order of ChPT.

The lowest cut in the $K_{\ell 3}$ form factors $f_+(t)$, $f_1(t)$ also occurs at
$t\ge(M_K+M_\pi)^2$, therefore the $K_{\ell 3}$ matrix element is real throughout phase space. 
In this work, we use a phenomenological parameterization of these form factors; 
see Appendix~\ref{app:Kl3formfactor}.

\subsection{$\boldsymbol{\Im{V_i}}$, $\boldsymbol{\Im{A_i}}$ at $\boldsymbol{\order(p^6)}$}

\begin{figure}
  \centering
  \epsfig{width=0.9\linewidth,file=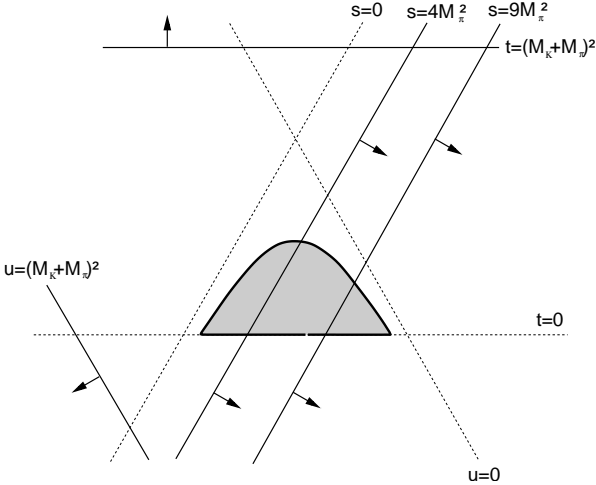}
  \caption{Mandelstam plane with the lowest cuts in the variables $s$, $t$, $u$, 
           at fixed $W^2=m_e^2$.}
\label{fig:Mandelstamplane}
\end{figure}
The above argument holds in fact for all cuts in the variables $t$, $u$, and $W^2$, 
at arbitrary orders in the chiral expansion:
because of strangeness conservation, all cuts in these variables lie at 
$t,\,u,\,W^2 \ge (M_K+M_\pi)^2$ and therefore outside the physical region.
Cuts within the physical region can only be due to two- or three-pion intermediate
states in the $s$-channel, which first arise at $\order(p^6)$ in ChPT.
All lowest cuts in the variables $s$, $t$, and $u$, and their positions
in the Mandelstam plane are depicted in Fig.~\ref{fig:Mandelstamplane}.

The Wess--Zumino--Witten Lagrangian contains $\pi\pi\pi\gamma$ vertices, 
therefore it is possible to have $s$-channel cuts of one-loop diagrams 
that will contribute to the axial tensor $A_{\mu\nu}$.
\begin{figure}
  \centering
  \epsfig{width=0.65\linewidth,file=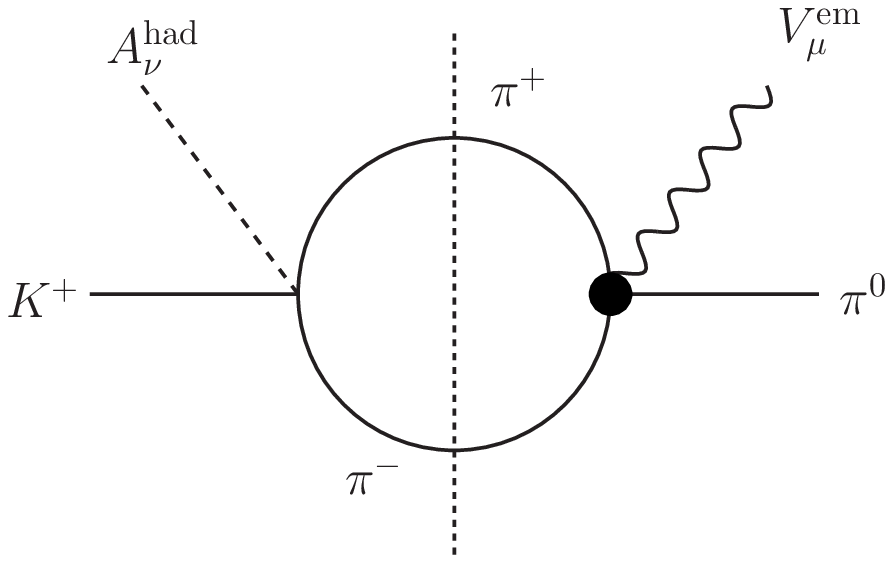}
  \epsfig{width=0.65\linewidth,file=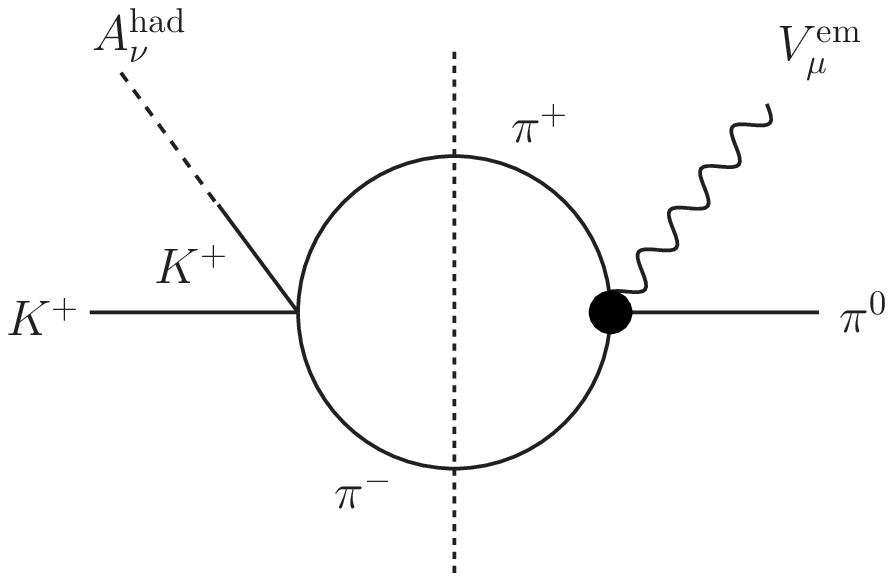}
  \caption{$\order(p^6)$ diagrams with cuts in the $s$-channel that contribute to the $\Im{A_i}$.
  The big filled dots denote vertices from the Wess--Zumino--Witten Lagrangian. 
  The second diagram with the kaon pole contributes to $A_3$ only.}
\label{fig:AcutOp6}
\end{figure}
The two diagrams shown in Fig.~\ref{fig:AcutOp6} are the only ones that have to be evaluated.
We find
\bea
   \Im{A_1} &=& \Im{A_3} = - \frac{s}{384\pi^3 F^4} \left(1-\frac{4M_\pi^2}{s}\right)^{3/2} ~, \notag\\
   \Im{A_2} &=& \Im{A_4} = 0 ~.
\label{eqn:summarizeA1loop}
\eea
The threshold behavior in \eqref{eqn:summarizeA1loop} is 
dictated by the fact that the intermediate pions are required to be in a relative $p$-wave,
which leads to a suppression by a factor of $(1-\nolinebreak4M_\pi^2/s) \sim |\vec{p}_{\operatorname{rel}}|^2$
(where $\vec{p}_{\operatorname{rel}}$ is the relative 3-momentum of the pions in their 
center-of-mass system).

At $\order(p^6)$, the vector correlator $V_{\mu\nu}$ can only receive three-pion-cut 
contributions in the physical region, i.e., from two-loop diagrams.
As the kinematical limit for $s$ in the physical region is
$s \le (M_K-m_\ell)^2$, these will only contribute in the electron channel.

\begin{figure}
  \centering
  \epsfig{file=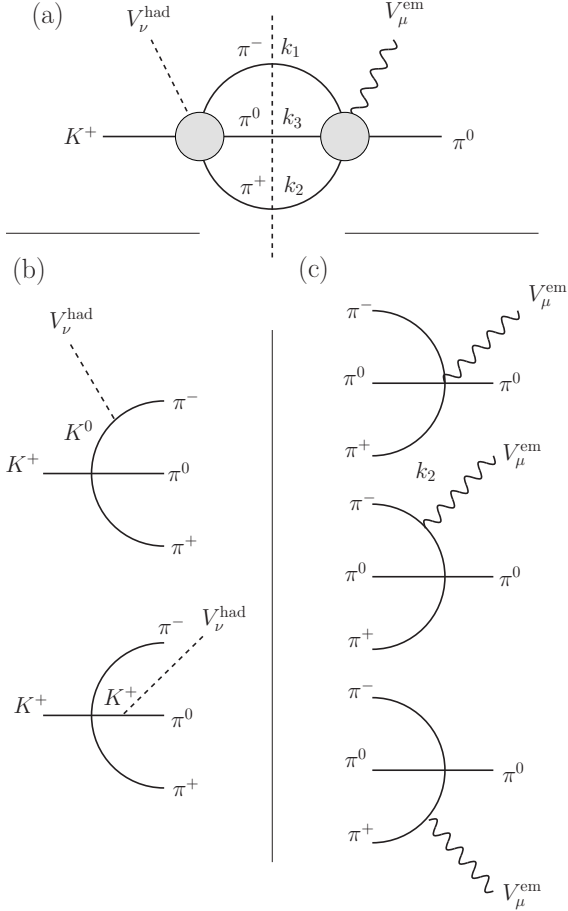,width=0.9\linewidth}
  \caption{Two-loop diagrams with three-pion cuts that contribute to the $\Im{V_i}$.
\label{fig:VcutOp6}
}
\end{figure}
The general form of these two-loop diagrams is shown in Fig.~\ref{fig:VcutOp6}a,
where the blobs denote
all possible combinations of propagators and vertices from $\lagr^{(2)}$.
The right blob is the sum of all possible ways of coupling the photon, 
which is shown in Fig.~\ref{fig:VcutOp6}c. 
One can show that, as the photon correlator $\Gamma_\mu$ is antisymmetric in $k_1 \leftrightarrow k_2$,
\bea
\Gamma_\mu &=& \langle\pi^0(p')\,|\,V_\mu^{\elm}\,|\,\pi^-(k_1)\pi^+(k_2)\pi^0(k_3)\rangle \\
&=& \frac{e}{F^2}\Bigl\{ (k_1+k_2-q)^2 -M_\pi^2 \Bigr\} 
\biggl( \frac{{k_1}_\mu}{k_1q}-\frac{{k_2}_\mu}{k_2q} \biggr) + \order(p^4)~, \notag
\eea
it is sufficient to consider the sum of diagrams shown in Fig.~\ref{fig:VcutOp6}b
for the weak vector current coupling, as the sum of all other diagrams does not contribute. 
Isospin symmetry conservation dictates that the correlator~\cite{Bla95}
\beq \begin{split}
  \langle\pi^-(k_1)  \pi^+(k_2) \pi^0(k_3)\,|\,V_\nu^{\had} \,|\, K^+(p)\rangle & \\
   =~ A_\nu(2,1,3) + B_\nu(2,1,3) & \label{eqn:Anu123}
\end{split} \eeq
can be decomposed into an amplitude $A_\nu(2,1,3)$ that is symmetric in the charged-pion momenta $k_1$, $k_2$,
and an amplitude $B_\nu(2,1,3)$ that is totally antisymmetric in $k_1$, $k_2$, and $k_3$.
Thus only $B_\nu(2,1,3)$
is projected out as the integration is symmetric in the intermediate pion momenta. Reversing this
argument it is also clear that only the totally antisymmetric part of the photon correlator $\Gamma_\mu$
will contribute. 
This leads to a further suppression of the imaginary parts 
by $1-9M_\pi^2/s$ near the cut threshold as we will discuss below.

In order to evaluate the imaginary part of these diagrams we use Cutkosky rules~\cite{Cut60}.
As the phase space is restricted by $s\le(M_K-m_e)^2$ and the cut starts at $s\ge9M_\pi^2$ the
intermediate pions are close to threshold. 
We thus expand the integral in the small quantity 
\begin{eqnarray}
  \epsilon = \frac{\sqrt{s}-3M_\pi}{M_\pi} = \frac{3}{2}\biggl(1-\frac{9M_\pi^2}{s}\biggr) + \order(\epsilon^2)~,
\end{eqnarray}
and content ourselves with calculating
the imaginary parts of the structure functions $V_i$ to leading order in $\epsilon$. 
In this way we correctly reproduce the leading threshold 
behavior.\footnote{Note that, despite the fact that $\epsilon$ can be as large as 0.5
at the border of phase space, such an expansion in $\epsilon$ is seen to converge rather
rapidly for the imaginary part of the scalar sunset diagram $S$, 
$\Im{S}=M_\pi^2 \epsilon^2/(384\sqrt{3}\pi^2) \{ 1-\epsilon/6+7\epsilon^2/144\mp\ldots\}$;
see e.g.\ Ref.~\cite{Gas98}.}

We first discuss how the threshold behavior of this three-pion-cut contribution
can be understood without explicit calculation, in terms of symmetry arguments.   
We argued above that only the totally antisymmetric part of the correlator 
in~\eqref{eqn:Anu123} will contribute. 
However, in our expansion $B_\nu(2,1,3)$ is expandable as a polynomial in the pion three momenta 
$\vec{k}_i$, and it is easy to see that each polynomial that is totally antisymmetric in the 
$\vec{k}_i$ is at least of $\order(\vec{k}_i\vec{k}_j) = \order(\epsilon)$. 
The same is true for the totally antisymmetric part of the photon 
correlator.
Note that both the kaon and the pion propagators in the diagrams Fig.~\ref{fig:VcutOp6}
are non-singular near threshold,
\bea
M_K^2 - (k_i+W)^2 & \rightarrow& \frac{2}{3}\bigl(M_K^2+3M_\pi^2-W^2\bigr) + \order(\sqrt{\epsilon}) ~,\notag\\
k_i q &\to& \frac{4}{3}M_\pi^2 + \order(\sqrt{\epsilon}) ~
\eea
(in the center-of-mass system of the three pions), and
can therefore be safely expanded.
We therefore obtain a factor of $\epsilon^2$ from the vertices alone. 
The three-particle phase space by itself is of $\order(\epsilon^2)$,
hence we expect that the lowest power of $\epsilon$ in 
the expansion of the imaginary parts of the structure functions $V_i$ will be 
$\epsilon^4 \sim (1-9M_\pi^2/s)^4$. 
So while angular momentum conservation leads to a suppression by one additional power
of $\epsilon$ like in the two-pion-cut diagrams discussed above, another factor
of $\epsilon$ is due to isospin symmetry arguments.  If isospin-violating effects 
were allowed for, an isospin-breaking threshold behavior $\sim \epsilon^3$ would occur.

\begin{sloppypar}
Further technical details on the calculation of the three-pion-cut contributions
can be found in Ref.~\cite{Mul06}.
The explicit expressions we find are 
\begin{eqnarray}
   \Im{V_i} &=& C_i\cdot\frac{1}{32768\pi^2\sqrt{3}F^4}  \label{eqn:2loopthreshold} \\
   &\times&\frac{s}{M_K^2+3M_\pi^2-W^2}
      \,\biggl(1-\frac{9M_\pi^2}{s}\biggr)^4  
   + \order\bigl(\epsilon^5\bigr) ~,\notag
\end{eqnarray}
with
\begin{xalignat}{2}
    C_1 &= \frac{1}{2}\bigl(s+t-M_K^2\bigr)~, &
    C_2 &= \frac{1}{2}\bigl(s+M_\pi^2\bigr) ~, \notag\\
    C_3 &= 0 ~,&
    C_4 &= 1 ~. \label{eqn:Ci}
\end{xalignat}

We expect that the two types of contributions to the $T$-odd asymmetry 
due to imaginary parts in the structure functions $A_i$, $V_i$ 
differ considerably in size for the following reasons:
\begin{enumerate}
  \item The three-pion-cut diagrams contribute only in a very small region of phase
    space compared to the two-pion-cut diagrams due to the higher threshold.
    Numerically, this turns out to account for a relative suppression by roughly three orders of magnitude.
  \item The threshold behavior of the latter is only suppressed as $(1-4M_\pi^2/s)^{3/2}$ compared to 
    $(1-9M_\pi^2/s)^4$ for the three-pion-cut diagrams.  The higher exponent for the three-pion cut
    leads to a further suppression by about two orders of magnitude.
  \item Finally, additional combinatorial prefactors of the three-particle 
    versus the two-particle phase space suppress the former by another two orders of magnitude.
\end{enumerate}
Altogether we expect the three-pion-cut diagrams at $\order(p^6)$ to be suppressed 
by roughly seven orders of magnitude 
compared to the two-pion-cut contributions at the same chiral order.
These expectations are born out in the precise numerical evaluations below.
\end{sloppypar}

\subsection{Beyond $\boldsymbol{\order(p^6)}$}

In the light of the strong kinematical suppression of the three-pion-cut contributions
in the vector structure functions, we find better constraints on the asymmetries
due to hadronic loops by calculating the leading chiral \emph{two-pion cuts} instead
of the leading chiral imaginary parts in general.  
Even though two-pion-cut contributions in the $V_i$ only occur at $\order(p^8)$
and therefore at higher chiral order than the three-pion-cut terms, 
larger phase space and less strongly suppressed threshold behavior will actually
make these contributions numerically much more significant.
In addition, for the muon channel these represent the first cuts in the vector structure functions
at all.\footnote{In Ref.~\cite{Gas05}, only the first cuts at $\order(p^6)$ were
considered, which is why the two-pion cut in the vector structure functions 
is not shown in Fig.~2 of that reference.}

\begin{figure}
  \centering
  \epsfig{file=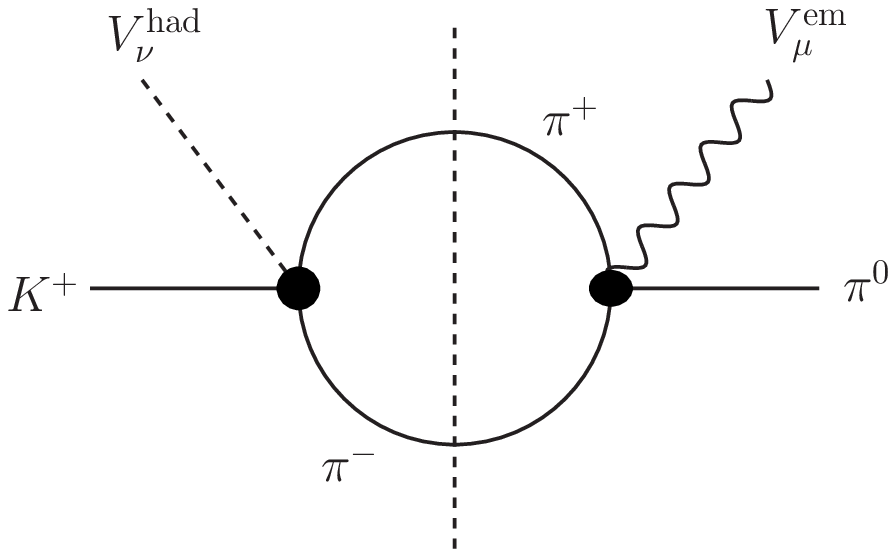,width=0.65\linewidth}
  \caption{Two-pion-cut diagram contributing to $\Im{V_i}$ at $\order(p^8)$.}
\label{fig:oneloopOp8}
\end{figure}
\begin{sloppypar}
At $\order(p^8)$, the one-loop diagram in Fig.~\ref{fig:oneloopOp8} with two anomalous vertices  
contributes to the imaginary parts of the $V_i$.
The diagram is readily evaluated with the result
\begin{eqnarray}
    \Im{V_i} &=& C_i\cdot \frac{s}{1536 \pi^5 F^6}\left(1-\frac{4M_\pi^2}{s}\right)^{3/2}
    \label{eqn:ImV1loop}
\end{eqnarray}
with the same prefactors $C_i$ of \eqref{eqn:Ci}. 
Note that, again, as the intermediate pions cannot be in a
relative $s$-wave the diagram is suppressed by an additional factor of 
$(1-\nolinebreak4M_\pi^2/s)$ compared to the scalar diagram.
We therefore find the same threshold behavior as for the axial vector structure functions.
\end{sloppypar}

While we have evaluated the leading chiral two-pion-cut contributions to 
$A_{\mu\nu}$ \eqref{eqn:summarizeA1loop}, and $V_{\mu\nu}$ \eqref{eqn:ImV1loop}, 
the structure functions $A_2$, $A_4$, and $V_3$ are still real at these orders.
We have checked by inserting polynomial vertices of higher order in 
the diagrams of Figs.~\ref{fig:AcutOp6} and \ref{fig:oneloopOp8} 
that this is in general not the case.  
As a complete calculation of such even-higher-order effects is beyond the
scope of this article, we only want to ensure that asymmetries due
to imaginary parts in $A_2$, $A_4$, and $V_3$ are not accidentally enhanced
by kinematical prefactors in the squared matrix element.  
To convince ourselves that higher order corrections are sufficiently suppressed, 
we will below evaluate the asymmetries with the imaginary parts of $A_1$
and $V_4$ inserted by hand into $A_2$, $A_4$, and $V_3$:
\begin{eqnarray}
    \Im{A_2} &=& M_K^2\cdot\Im{A_4} 
    = \frac{s}{384\pi^3 F^4}\left(1-\frac{4M_\pi^2}{s}\right)^{3/2} ~,\notag \\[2ex] 
    \Im{V_3} &=& \frac{s}{1536\pi^5 F^6}\left(1-\frac{4M_\pi^2}{s}\right)^{3/2} ~.
    \label{eqn:pseudoIm} 
\end{eqnarray}
In reality, $\Im{V_3}$, $\Im{A_2}$, $\Im{A_4}$ are suppressed
compared to \eqref{eqn:pseudoIm} by at least two orders in the chiral expansion.
The ``pseudo-asymmetries'' calculated from \eqref{eqn:pseudoIm} are therefore
conservative upper-limit estimates for such effects.

\subsection{Numerical results}

We compute the differential decay width $d\Gamma/d\xi$ in the rest frame of the decaying kaon 
by evaluating the phase space integral 
using the phase space generator \texttt{RAMBO}~\cite{Kle85}.
In the electron channel we apply the ``standard'' cuts on the photon energy ($\Eg>30$~MeV)
and on the photon positron angle ($\te>20\degree$), whereas in the muon channel we only 
cut on the photon energy.

\begin{table}
\caption{Asymmetries in the electron- and muon channel. $V_i$ ($A_i$) labels the asymmetry 
we would obtain if we set all imaginary parts except $\Im{V_i}$ ($\Im{A_i}$) to zero.}
\label{tab:asymmetrynumerics}
\centering
\renewcommand{\arraystretch}{1.2}
\begin{tabular}{cccc}
  \hline\noalign{\smallskip}
         & & electron & muon \\
\noalign{\smallskip}\hline\noalign{\smallskip}
         & $A_1$ & $1.4\tee{-6}$ & $4.5\tee{-7}$ \\ 
         & $A_2$ & $0$           & $0$           \\ 
         & $A_3$ & ---           & $-0.04\tee{-7}$ \\ 
         & $A_4$ & $0$           & $0$           \\ 
\noalign{\smallskip}\hline\noalign{\smallskip}
         & $V_1$ & $-1.2\tee{-7}$ & $-3.1\tee{-8}$ \\ 
         & $V_2$ & $-0.9\tee{-7}$ & $-4.2\tee{-8}$ \\ 
         & $V_3$ & ---            & $ 0          $ \\ 
\oneloop & $V_4$ & $-2.2\tee{-7}$ & $-0.2\tee{-8}$ \\
\noalign{\smallskip}\hline\noalign{\smallskip}
         & $V_1$ & $-0.8\tee{-14}$ &  --- \\ 
         & $V_2$ & $-3.0\tee{-14}$      &  --- \\ 
         & $V_3$ & ---                  & ---  \\ 
\twoloop & $V_4$ & $-6.6\tee{-14}$ & --- \\
\noalign{\smallskip}\hline\noalign{\smallskip}
         total & & $0.9\tee{-6}$ & $3.7\tee{-7}$ \\
\noalign{\smallskip}\hline
\end{tabular}
\renewcommand{\arraystretch}{1.0}
\end{table}
\begin{sloppypar}
The numerical results for the contributions to the asymmetry $A_\xi$ from the various loop diagrams 
are shown in Table~\ref{tab:asymmetrynumerics}. 
Note that the imaginary parts of the $A_i$ yield the dominant contribution. 
The main contribution to the $\Im{V_i}$ is of $\order(p^8)$, 
and the numerical suppression relative to the $\order(p^6)$
$\Im{A_i}$ terms indeed turns out to be about one order of magnitude.
The asymmetry in the muon channel is slightly smaller due to 
relatively smaller phase space available for imaginary parts from the two-pion cut.
We point out once more that due to the necessary $\pi\pi\to\pi\gamma$ rescattering,
all these two-pion cut contributions can be traced back to the axial anomaly.
The relative size of the two-loop contributions agrees roughly with 
our estimate in the previous section, i.e.\ they are suppressed by seven orders of magnitude
and are totally negligible.
\end{sloppypar}

\begin{table}
\caption{Asymmetries due to non-vanishing imaginary parts in $V_3$, $A_2$, $A_4$;
see \eqref{eqn:pseudoIm}.}
\label{tab:pseudoasymmetry}
\centering
\renewcommand{\arraystretch}{1.2}
\begin{tabular}{ccc}
  \hline\noalign{\smallskip}
         & electron & muon \\
\noalign{\smallskip}\hline\noalign{\smallskip}
         $V_3$ & ---           & $ 1.4\tee{-8}$ \\ 
\noalign{\smallskip}\hline\noalign{\smallskip}
         $A_2$ & $-0.8\tee{-6}$ & $-4.5\tee{-7}$ \\ 
         $A_4$ & $ 0.5\tee{-6}$ & $ 0.5\tee{-7}$ \\ 
\noalign{\smallskip}\hline
\end{tabular}
\renewcommand{\arraystretch}{1.0}
\end{table}
Finally, we quote the pseudo-asymmetries 
estimated from the choice of imaginary parts given in \eqref{eqn:pseudoIm}
in Table~\ref{tab:pseudoasymmetry}.
We find all of them to be at most of the same order as the 
non-vanishing contributions in the other structure functions, 
hence there is no unnatural kinematical enhancement of 
$\Im{V_3}$, $\Im{A_2}$, and $\Im{A_4}$, such that an estimate
of typical chiral SU(3) higher-order corrections of the order of 20--30\%
remains valid.

\begin{figure}
  \centering
  \epsfig{file=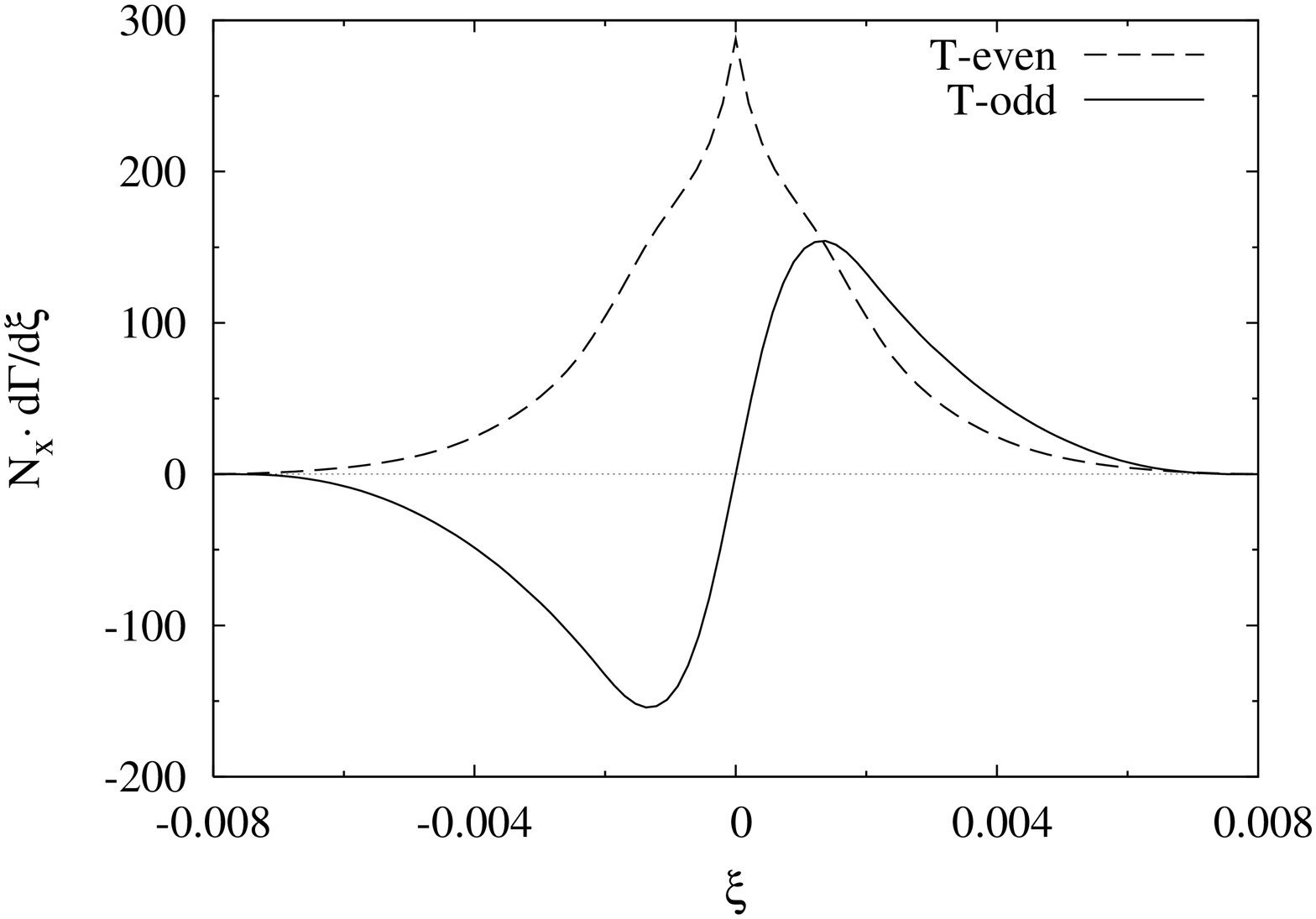,angle=270,width=0.95\linewidth}
  \caption{$T$-even and $T$-odd contributions to $d\Gamma/d\xi$ in the electron channel. 
     $10^6N_{\odd}^{-1} = N_{\even}^{-1} = \Gamma(K_{e3\gamma})$}
\label{fig:plotxielectron}
  \centering
  \epsfig{file=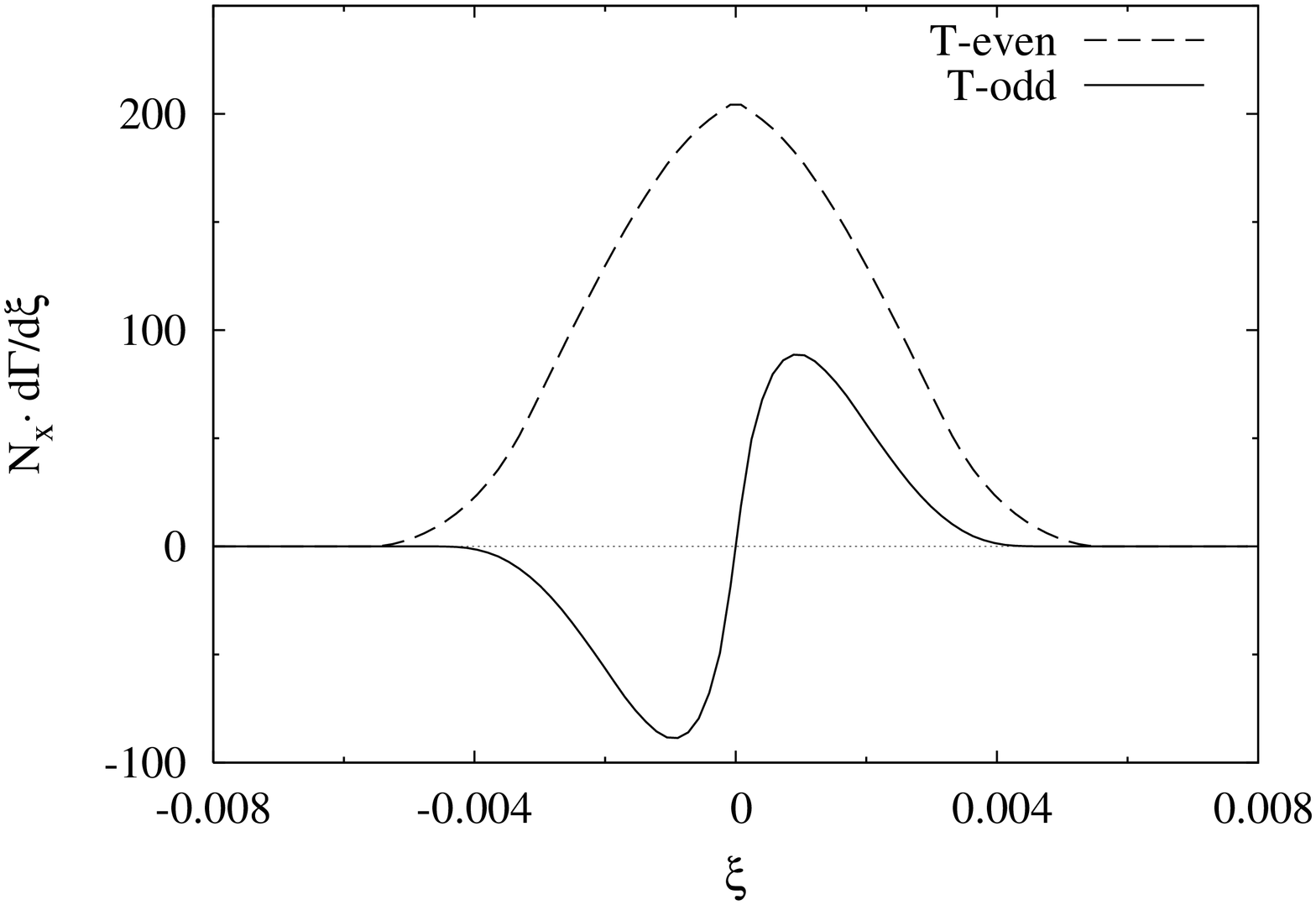,angle=270,width=0.95\linewidth}
  \caption{$T$-even and $T$-odd contributions to $d\Gamma/d\xi$ in the muon channel. 
     $10^6N_{\odd}^{-1} = N_{\even}^{-1} = \Gamma(K_{\mu 3\gamma})$}
\label{fig:plotximuon}
\end{figure}

In Figs.~\ref{fig:plotxielectron} and \ref{fig:plotximuon} 
we also plot the $T$-even and $T$-odd contributions to the differential width $d\Gamma/d\xi$,
both for the electron and the muon channel.
As $\xi^2$ can be expressed in terms of the standard five invariant kinematical variables, 
the distribution $d\Gamma/d\xi$ contains no fundamentally new information beyond 
differential widths with respect to those other variables plus the integral asymmetry
$A_\xi$.

\section{Comparison to other $\boldsymbol{T}$-odd contributions}\label{sec:comparison}

\subsection{$\boldsymbol{T}$-odd correlations and photon loops}

\begin{figure}
  \centering
  \epsfig{file=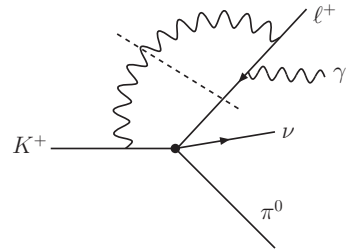,width=0.6\linewidth}
  \caption{Example of a photon loop diagram. 
  For a complete listing of all Feynman diagrams that contribute to the
  imaginary parts via electromagnetic final-state interactions, see Fig.~5 in Ref.~\cite{Bra02}.}
\label{fig:photonloop}
\end{figure}

In the previous section, we evaluated the correlators $V_{\mu\nu}$ and $A_{\mu\nu}$ in ChPT 
and showed that the structure functions $V_i$, $A_i$ become complex at higher orders in the chiral expansion,  
which leads to a non-vanishing value of the asymmetry $A_\xi$.
However, the matrix element \eqref{eqn:Tsquared} we use is only of lowest order in $\alpha$.
In Ref.~\cite{Bra02}, higher-order electromagnetic corrections to this matrix
element were included. These consist of loop diagrams with intermediate photons as the one in 
Fig.~\ref{fig:photonloop} that develops an imaginary part. Squaring the amplitude 
(which, in addition to \eqref{eqn:Tsquared}, contains next-to-leading-order electromagnetic 
corrections denoted $T_{\operatorname{one\ loop}}$ in (10) in Ref.~\cite{Bra02}) 
one sees that these imaginary parts lead to a non-zero $A_\xi$.
The following asymmetries in the electron and in the muon channel were obtained~\cite{Bra02}: 
\beq \begin{split}
  A_\xi\bigl(K^+\rightarrow\pi^0  e^+\nu_e  \gamma\bigr) &= -0.59\tee{-4}~,\\
  A_\xi\bigl(K^+\rightarrow\pi^0\mu^+\nu_\mu\gamma\bigr) &=  1.14\tee{-4}~,
\end{split}\eeq
which have to be compared to the values we get from hadronic loops 
(Table~\ref{tab:asymmetrynumerics}).\footnote{As noted in Ref.~\cite{Bra02},
the photon loop asymmetry is larger (and opposite in sign) in the muon channel due
to numerically significant structures scaling with the lepton mass.}
We find that the effects of strong interaction phases are suppressed 
compared to the electromagnetic final-state interaction by roughly two orders of magnitude.

\begin{sloppypar}
Naively one might have expected that both are of comparable size because, 
although the electromagnetic asymmetry is suppressed by $\alpha \sim 1/137$, 
hadronic loops only occur in the SD part of the matrix element, 
which is suppressed by a factor of $\sim 1/100$. 
However, the following two additional effects discussed in the previous section
reduce the impact of the imaginary parts from pion loops even further 
compared to the photon loop contributions:
\begin{enumerate}
\item While the imaginary parts of photon loop diagrams are non-vanishing
in the complete physical phase space, those from pion loops only
contribute to the asymmetry $A_\xi$ above the two-pion threshold, $s>4M_\pi^2$.
This reduced part of phase space accounts for a suppression of roughly
one order of magnitude. 
\item In addition, the slow rise of the imaginary part above threshold
due to its $p$-wave characteristic reduces its effect even further.
\end{enumerate}
Altogether this explains the relative smallness of the hadronic asymmetries.
The standard model contributions to the $T$-odd correlations in $\kl3g$
are therefore indeed dominated by electromagnetic final-state interactions,
and the estimate given in Ref.~\cite{Bra02} remains valid.
Hadronic loops are suppressed even to the extent that the main uncertainty
are effects of chiral order $\order(e^2p^4)$ that have not yet been considered.
\end{sloppypar}

\subsection{$\boldsymbol{T}$-odd correlations beyond the standard model}\label{sec:bySM}

\begin{sloppypar}
So far, we only considered contributions to the asymmetry $A_\xi$ from
pion and photon loop effects.  We now turn to $T$-odd correlations
due to generalized weak current-current interactions,
which can arise in certain extensions of the standard model.
In Refs.~\cite{Bra03,Ger04} the following model-independent
effective Lagrangian for $s\rightarrow u$ transitions is given:
  \begin{eqnarray}
  \lagr &=& \frac{G_F}{\sqrt{2}}V_{us}^*\Big[
  \sbar\gamma^\mu\bigl((1+g_v)-(1-g_a)\gamma_5\bigr)u \cdot \nubar (1+\gamma_5)\gamma_\mu\ell  \notag\\
  &&\qquad+\; \sbar \bigl(g_s + g_p \gamma_5\bigr)u \cdot \nubar           (1+\gamma_5)\ell \notag\\
  &&\qquad+\; g_t\,\sbar\sigma^{\mu\nu}    u \cdot \nubar (1+\gamma_5)\sigma_{\mu\nu}\ell 
  \Big] + (h.c.) ~, \label{eqn:BSMlagrangian}
\end{eqnarray}
with  $\sigma_{\mu\nu} = \frac{i}{2}[\gamma_\mu,\gamma_\nu]$.
In contrast to the standard model, which only contains weak $V-A$ 
transitions, there are also terms that allow for transitions via purely vector,
axial vector, scalar, pseudoscalar, and tensor currents. 
For a discussion of models generating such effective couplings 
and possible constraints on their size from $\kl3g$, 
see Ref.~\cite{Bra03} and references therein.
\end{sloppypar}

From the Lagrangian \eqref{eqn:BSMlagrangian} one can immediately obtain
a generalized matrix element for the $\kl3g^+$ decay
\begin{eqnarray}
  T &=& \frac{G_F}{\sqrt{2}}V_{us}^*e\,\epsilon^\mu(q)^*\notag\\
  &\times&\Big[
  \big((1+g_v)V_{\mu\nu}-(1-g_a)A_{\mu\nu}\big)\cdot\nubar(1+\gamma_5)\gamma^\nu\ell
  \notag\\
  &+&(1+g_v)\frac{F_\nu}{2p_\ell q}\cdot\nubar(1+\gamma_5)\gamma^\nu
  (m_\ell-\peslash-\qslash)\gamma_\mu\ell\notag\\
  &+&\bigl(g_s \,F^s_\mu+g_p \,F^p_\mu\bigr)\cdot \nubar(1+\gamma_5)\ell\notag\\
  &+&g_s\,\frac{f}{2p_\ell q}\cdot \nubar(1+\gamma_5)
  (m_\ell-\peslash-\qslash)\gamma_\mu\ell\notag\\
  &+& g_t \,\hat{T}_{\mu\nu\rho} \cdot \nubar(1+\gamma_5)\sigma^{\nu\rho}\ell\notag\\
  &+& g_t \,\frac{F^t_{\nu\rho}}{2p_\ell q} \cdot \nubar(1+\gamma_5)\sigma^{\nu\rho}
  (m_\ell-\peslash-\qslash)\gamma_\mu\ell
  \Big] ~, \label{eqn:TBSM}
\end{eqnarray}
where the hadronic correlators are defined by\footnote{Note that the scalar, pseudoscalar,
and tensor correlators are not renormalization-group invariant and have to be
taken at a specific scale.  
If one calculates the coupling constants $g_s$, $g_p$, $g_t$ in a certain model,
they have to be chosen accordingly.
As a consequence of this, we have to specify quark mass values for an evaluation
of the scalar and pseudoscalar correlators, see Appendix~\ref{app:spDeriv}, 
which are scale-dependent.\label{foot:scale}}
\newcommand{\ddx}{\mathcal{D}^4x}
\begin{eqnarray}
  V_{\mu\nu} &=& i\int \ddx \langle\pi^0(p')|T\bigl\{V^{\elm}_\mu(x)(\sbar\gamma_\nu u)(0)\bigr\}|K^+(p)\rangle ~,\notag\\
  A_{\mu\nu} &=& i\int \ddx \langle\pi^0(p')|T\bigl\{V^{\elm}_\mu(x)(\sbar\gamma_\nu\gamma_5 u)(0)\bigr\}|K^+(p)\rangle ~,\notag\\
  F_\nu &=& \langle\pi^0(p')|(\sbar\gamma_\nu u)(0)|K^+(p)\rangle ~,\notag\\
  F^s_\mu &=& i\int \ddx \langle\pi^0(p')|T\bigl\{V^{\elm}_\mu(x)(\sbar u)(0)\bigr\}|K^+(p)\rangle ~,\notag\\
  F^p_\mu &=& i\int \ddx \langle\pi^0(p')|T\bigl\{V^{\elm}_\mu(x)(\sbar \gamma_5  u)(0)\bigr\}|K^+(p)\rangle ~,\notag\\
  f &=& \langle\pi^0(p')|(\sbar u)(0)|K^+(p)\rangle ~,\notag\\
  \hat{T}_{\mu\nu\rho} &=& i\int \ddx
  \langle\pi^0(p')|T\bigl\{V^{\elm}_\mu(x) (\sbar\sigma_{\nu\rho}u)(0)\bigr\}|K^+(p)\rangle ~,\notag\\
  F^t_{\nu\rho} &=& \langle\pi^0(p')|(\sbar\sigma_{\nu\rho}u)(0)|K^+(p)\rangle ~,\label{eqn:BSMformfactors}
\end{eqnarray}
with
\begin{eqnarray}
  \int \ddx &\equiv& \int d^4x \,e^{iqx} ~.
\end{eqnarray}
Squaring the matrix element \eqref{eqn:TBSM} one finds that the $T$-odd part is proportional to 
the imaginary parts of the couplings $g_i$
\begin{eqnarray}
  \sN^{-1}\sum_{\operatorname{spins}} |T|^2 
    &=& |T|^2_{\even} 
    + \xi \Bigl(\Im{g_v} C_v + \Im{g_a} C_a  \label{eqn:TsquaredBSM}  \\
    && +\Im{g_s}C_s  + \Im{g_p}C_p + \Im{g_t}C_t\Bigr) ~,~\notag
\end{eqnarray}
where the $C_i$ are functions depending on the $K_{\ell 3}$ form factors and the
vector and axial structure functions.
One can compute the differential width $d\Gamma/d\xi$ by integrating over phase space 
and extract the various parts $A_\xi^{(i)}$ of the asymmetry $A_\xi$
that we decompose according to
\beq
A_\xi ~= \sum_{i=v,a,s,p,t}  \Im{g_i} \, A_\xi^{(i)} ~.
\eeq
The numerical results for the $A_\xi^{(i)}$ are shown in Table~\ref{tab:BSMasymmetries}.
They are a measure for the sensitivity of an asymmetry determination to
the imaginary parts of the different couplings $g_i$.
We note that the numbers for scalar and pseudoscalar asymmetries 
disagree somewhat with the precise results quoted in Ref.~\cite{Bra03}.
\begin{table}
\caption{Asymmetries $A_\xi^{(i)}$ for extensions of the standard model,
where $i$ is listed in the first column.}
\label{tab:BSMasymmetries}
\centering
\renewcommand{\arraystretch}{1.2}
\begin{tabular}{ccc}
  \noalign{\smallskip}\hline\noalign{\smallskip}
          & electron & muon\\
  \noalign{\smallskip}\hline\noalign{\smallskip}
          vector/axial vector & $-2.5\tee{-3}$ & $-7.8\tee{-3}$ \\
          scalar             & $-3.6\tee{-5}$ & $-1.3\tee{-2}$ \\
          pseudoscalar       & $-1.4\tee{-5}$ & $-4.8\tee{-3}$ \\
          tensor             & $ 1.7\tee{-5}$ & $ 6.0\tee{-3}$ \\
  \noalign{\smallskip}\hline\noalign{\smallskip}
\end{tabular}
\renewcommand{\arraystretch}{1.0}
\end{table}

We want to shed some additional light on the structure of the terms $C_i$
and the numerical size of the $A_\xi^{(i)}$ as displayed in Table~\ref{tab:BSMasymmetries}.
As the infrared divergences for small photon momenta in $\kl3g$ are intimately
linked to virtual photon corrections in the non-radiative process $K_{\ell 3}$, 
it is obvious that the $T$-odd numerator in the asymmetry~\eqref{eqn:defAxi}
cannot be infrared divergent, while the denominator is.
In this sense, we would expect the asymmetries $A_\xi^{(i)}$
to be roughly of the order of the relative size SD/IB, that is on the percent level.
This proves to be a realistic estimate, see Table~\ref{tab:BSMasymmetries}.
Indeed, some of the asymmetries are proportional to structure dependent terms
($A_\xi^{(v/a)}$, see below, and $A_\xi^{(p)}$), while for those in which
interference terms between different bremsstrahlung contributions occur
($A_\xi^{(s)}$, $A_\xi^{(t)}$), the kinematic prefactors cancel out the 
appropriate power of photon momenta.

The structures $C_s$, $C_p$, and $C_t$ are multiplied by a factor of $m_\ell$ 
(see \eqref{eqn:VmunuFSmu}, \eqref{eqn:FpmuP}, and \eqref{eqn:Thatmunurho}),
which is why these asymmetries are suppressed in the electron channel 
compared to the ones proportional to $\Im{g_v}$ / $\Im{g_a}$.

It is argued in Ref.~\cite{Bra03} that $C_a=C_v$, as an overall phase
compared to the standard model $V-A$ coupling would not lead to any asymmetry.
As $g_a$ can obviously not contribute to the squared matrix element
if $A_{\mu\nu}$ vanishes, we conclude that $C_a=C_v$ has to be 
proportional to the axial vector structure functions $A_i$. 
Indeed we find the following decomposition of the asymmetries $A_\xi^{(v/a)}$
(for approximately constant structure functions)
\begin{align}
A_\xi^{(v/a)}&\bigl(K^+\to\pi^0 e^+\nu_e\gamma\bigr)  \notag\\
&= \bigl[ 1.8\av{A_1} - 1.5\av{A_2} + 0.5\av{A_4}\bigr] \cdot 10^{-3} ~, \notag\\
A_\xi^{(v/a)}&\bigl(K^+\to\pi^0 \mu^+\nu_\mu\gamma\bigr) \label{eqn:AvaDecomp} \\
&=  \bigl[ 5.1\av{A_1} - 5.3\av{A_2} + 0.2\av{A_3} + 0.8\av{A_4}\bigr] \cdot 10^{-3} ~. \notag
\end{align}
The total values in Table~\ref{tab:BSMasymmetries} are obtained with 
the values for the $\av{A_i}$ given in Table~\ref{tab:Vcentralvalues}.
It is obvious that chiral higher-order corrections modifying the $\av{A_i}$ 
will affect the asymmetries $A_\xi^{(v/a)}$ accordingly.
(This explains a slight deviation from the results for the vector/axial vector
asymmetries in Ref.~\cite{Bra03}, as our $\av{A_i}$ in Table~\ref{tab:Vcentralvalues} 
are smaller by a factor of $F_K/F_\pi \approx 1.2$.)
We emphasize that only the presence of the axial structure functions
due to the Wess--Zumino--Witten anomaly allows for a $T$-odd asymmetry
that scales with $\Im{g_v}$ / $\Im{g_a}$.
Furthermore, it follows that, as a matter of principle, $\Im{g_v}$ / $\Im{g_a}$
can only be extracted from experiment with the same accuracy with which
the axial structure functions are known.

We also split the form factors and structure functions for scalar and 
tensor interactions in \eqref{eqn:BSMformfactors} into an 
inner bremsstrahlung and a structure dependent part 
(see Appendix~\ref{app:BSMformfactors}, compare Sect.~\ref{sec:matrixelement}). 
We note again that interference terms of doubly structure dependent
origin are strongly suppressed and can be neglected in all cases.
For the part of the asymmetry proportional to $\Im{g_s}$,
a  decomposition analogous to \eqref{eqn:AvaDecomp}
is of the following form:
\begin{align}
A_\xi^{(s)}&\bigl(K^+\to\pi^0 e^+\nu_e\gamma\bigr) \notag\\
&=  \bigl[ 0.9 + 0.6 \av{S} - 0.2 \av{V_3} - 0.2 \av{V_4} \notag \\
& \quad+ 3.4 \av{A_1} - 0.5 \av{A_2} + 0.4 \av{A_3} + 0.7 \av{A_4} \bigr] \cdot 10^{-5} ~,
\notag \\ 
A_\xi^{(s)}&\bigl(K^+\to\pi^0 \mu^+\nu_\mu\gamma\bigr) \label{eqn:AsDecomp}\\
&=   \bigl[ 3.5 + 1.8 \av{S} - 0.6 \av{V_3} - 0.5 \av{V_4}  \notag\\
& \quad + 10.7 \av{A_1} - 4.7 \av{A_2} + 2.1 \av{A_3} + 1.8 \av{A_4}\bigr] \cdot 10^{-3} ~.
\notag 
\end{align}
(For a definition of the scalar structure function $S$, see Appendix~\ref{app:stIBSD}.)
Here, interference between vector and scalar bremsstrahlung terms occurs,
so there is a term that is not proportional to any of the structure functions,
but it turns out to be small, so that in particular the contribution
$\propto \av{A_1}$ yields the largest part of this asymmetry.  
So also the asymmetry $A_\xi^{(s)}$ is largely due to the existence
of the anomaly term.

\begin{sloppypar}
The pseudoscalar form factor $F_\mu^p$ is itself a purely structure dependent
term~\cite{Bra03}, see Appendix~\ref{app:spDeriv}, that can be calculated
at leading order from the Wess--Zumino--Witten anomaly.  
\end{sloppypar}

Only for the tensor asymmetry $A_\xi^{(t)}$, interference terms of 
purely bremsstrahlung type are dominant and by themselves lead to the
numbers quoted in Table~\ref{tab:BSMasymmetries} at a few percent accuracy.
Here, the vector and axial vector structure functions play no important role.
(We have disregarded structure dependent tensor contributions, about which
nothing is known; see Appendix~\ref{app:stIBSD}.)

\subsection{Cut dependence}\label{sec:cutdep}

In this section, we want to briefly investigate how the various
contributions to the asymmetries scale with the 
cuts on $\Eg$ and (in the electron channel) $\te$.

\begin{figure}
  \centering
  \epsfig{file=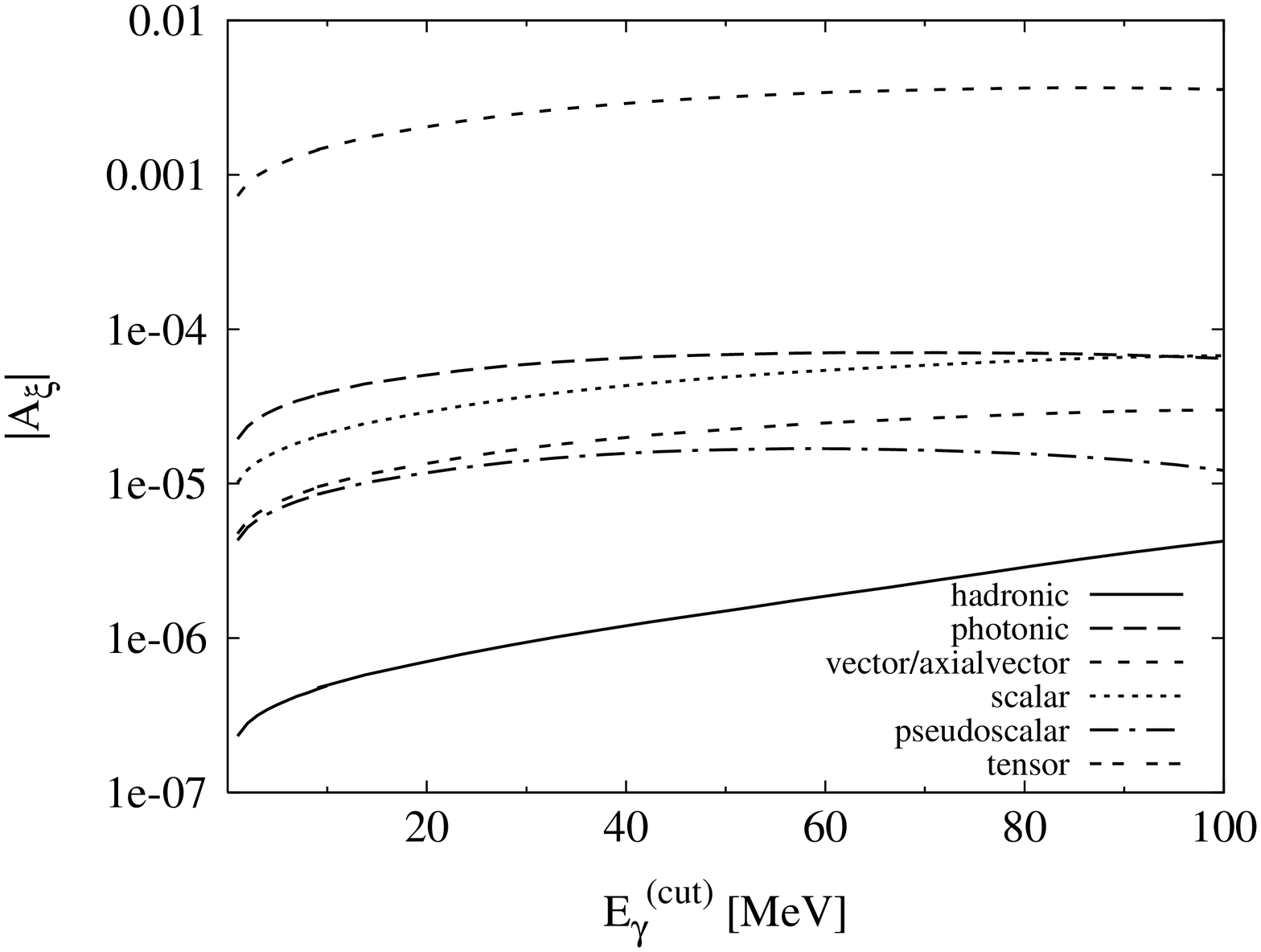,angle=270,width=\linewidth}
  \caption{Dependence of the various contributions to the $T$-odd asymmetry
           on $\Ecut$, for the electron channel, at fixed $\tecut=20\degree$.}
\label{fig:varycut_electron}
\end{figure}
\begin{figure}
  \centering
  \epsfig{file=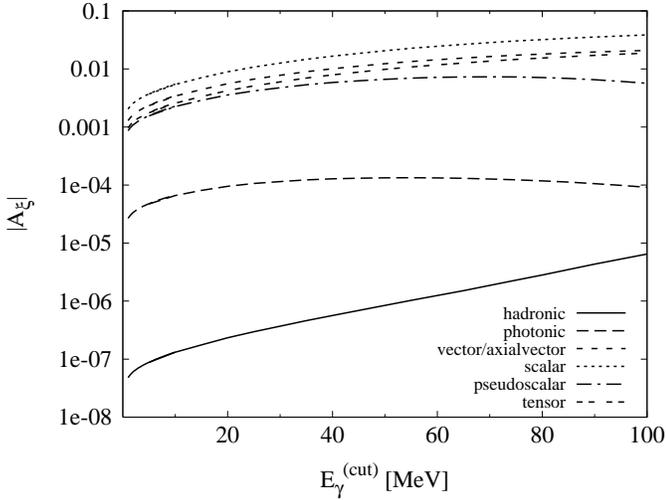,angle=270,width=\linewidth}
  \caption{Dependence of the various contributions to the $T$-odd asymmetry
           on $\Ecut$, for the muon channel.}
\label{fig:varycut_muon}
\end{figure}
\begin{figure}
  \centering
  \epsfig{file=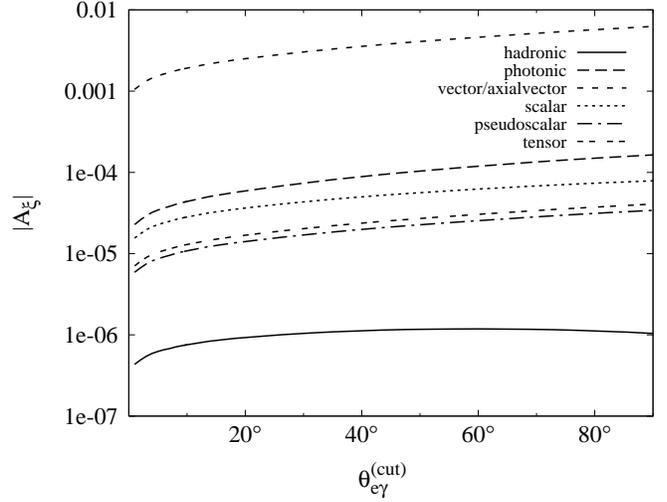,angle=270,width=\linewidth}
  \caption{Dependence of the various contributions to the $T$-odd asymmetry
           on $\tecut$, for the electron channel, at fixed $\Ecut=30$~MeV.}
\label{fig:varycut_angle}
\end{figure}
In Figs.~\ref{fig:varycut_electron} and \ref{fig:varycut_muon},
we show the $\Ecut$-dependence of the $T$-odd asymmetries due to 
pion loops, photon loops, as well as the $A_\xi^{(i)}$, $i=v/a,\,s,\,p,\,t$.  
Because of the infrared singularity in the 
$T$-even decay width, the asymmetries display a weak
singular behavior $\propto (\log \Ecut)^{-1}$ at very small energies;
Figs.~\ref{fig:varycut_electron} and \ref{fig:varycut_muon} show
that this singularity  is largely irrelevant for cuts
larger than 5~MeV.
The pion-loop contributions display the strongest rise with $\Ecut$
as the phase space regions with small $\Eg$ are those where
the imaginary parts of the hadronic structure functions vanish anyway.
Otherwise, there are no dramatic changes in the relative importance
of the different contributions in either channel,
such that the available statistics should determine the preferred
experimental cuts.

Figure~\ref{fig:varycut_angle} shows the $\tecut$-dependence
of the various contributions to the asymmetry for the electron channel.  
Here the variation of the hadronic loop contributions is particularly small.
A significant reduction of the asymmetries due to the near-singular behavior
of the denominator for collinear photon--electron momenta is only visible
for cuts well below $10\degree$.  Otherwise the conclusion is similar to the one for 
the photon energy cut dependence:  the relative importance of the various 
contributions hardly varies.

\section{Summary and conclusions}\label{sec:summ}

In this article, we have completed and deepened previous investigations of
$T$-odd correlations in radiative $K_{\ell 3}$ decays.  
We can summarize our findings as follows:
\begin{enumerate}
\item 
Although the vector and axial vector structure functions $V_i$, $A_i$
are real at chiral $\order(p^4)$, they develop imaginary parts stemming
from intermediate two- and three-pion states at higher orders. 
Starting at $\order(p^6)$, we have calculated the asymmetry
due to two-pion cuts in the $A_i$ and three-pion cuts in the $V_i$,
where the latter were found to be totally negligible due to phase space
and threshold behavior suppression.  
\item We have supplemented the leading imaginary parts in chiral power counting
by the leading chiral two-pion cuts, which only appear at $\order(p^8)$
in the case of the $V_i$. 
Altogether, this provides a reliable estimate of the 
asymmetries due to strong interaction phases, which we find to be 
$A_\xi = 0.9\times 10^{-6}$ for the electron channel and
$A_\xi = 3.7\times 10^{-7}$ for the muon channel.
Even-higher-order corrections to these numbers are not expected to exceed 20--30\%.
\item
These asymmetries turn out to be smaller by about two orders of magnitude
than those calculated from electromagnetic final-state interactions~\cite{Bra02}.
This suppression can be understood by the smallness of structure dependent
contributions in this decay channel, higher thresholds of the hadronic
intermediate states, and $p$-wave threshold behavior.  Altogether,
these effects overcompensate for the suppression of photon loops by 
the fine-structure constant $\alpha$.
We conclude that the estimate of the standard model contribution to 
the $T$-odd asymmetry based solely on photon loops remains valid.
\item 
We have re-analyzed the structure of the asymmetries that emerge due
to non-standard current-current interactions with complex coupling constants.
The generic size of such asymmetries at or below the percent level
can be understood in terms of the relative size of structure dependent terms
relative to the $T$-even bremsstrahlung amplitude.
We emphasize that the appearance of asymmetries proportional to imaginary
parts in vector, axial vector, scalar, and pseudoscalar coupling constants 
are due to the presence of the Wess--Zumino--Witten anomaly.
\item
We have finally shown that the relative importance of various contributions
to $T$-odd asymmetries show no strong dependence on the experimental cuts
$\Ecut$, $\tecut$.
\end{enumerate}
We conclude that the theoretical framework for an interpretation of 
$T$-odd asymmetries in $\kl3g^+$ decays is firmly set, and it remains 
a formidable, but potentially rewarding challenge to existing and future
experiments~\cite{Tch05,Bol05,Soz03,Shi05}
to measure such asymmetries at the required accuracy.

\begin{acknowledgement}
\textit{Acknowledgements.} 

\begin{sloppypar}  
We are grateful for the stimulating discussions with J.~Gasser, N.~Paver,
and G.~Colangelo that initiated this project.
Furthermore we thank V.~V.~Braguta and M.~Gerber for useful e-mail communications.
Partial financial support under the EU Integrated Infrastructure
Initiative Hadron Physics Project (contract number RII3-CT-2004-506078)
and DFG (SFB/TR 16, ``Subnuclear Structure of Matter'') is gratefully
acknowledged.
E.~H.~M. thanks the `Studienstiftung des deutschen Volkes' for supporting his studies.
\end{sloppypar}  
\end{acknowledgement}


\begin{appendix}
\suppressfloats
\renewcommand{\theequation}{\Alph{section}.\arabic{equation}}
\setcounter{equation}{0}

\section{Numerical parameters}

\subsection{Masses and decay constants}

In this article, we use the particle masses 
$M_K = M_{K^+} = 493.68$~MeV, 
$M_\pi = M_{\pi^0} = 134.98$~MeV,
$m_e = 0.511$~MeV,
$m_\mu = 105.658$~MeV.
In general, we work in the isospin limit.
We only adjust the thresholds in the loop functions such that the cuts in $s$
start in the appropriate place,
i.e.\ in \eqref{eqn:summarizeA1loop}, \eqref{eqn:2loopthreshold}, and \eqref{eqn:ImV1loop} 
we replace
\beq
  \biggl(1-\frac{(2M_\pi)^2}{s}\biggr)^{3/2} ~\longrightarrow~
  \biggl(1-\frac{(2M_{\pi^\pm})^2}{s}\biggr)^{3/2}
\eeq
for the two-pion cuts and
\beq
  \biggl(1-\frac{(3M_\pi)^2}{s}\biggr)^4  ~\longrightarrow~
  \biggl(1-\frac{(2M_{\pi^\pm}+M_{\pi^0})^2}{s}\biggr)^4
\eeq
for the three-pion cut contributions, where we use $M_{\pi^\pm}=139.57$~MeV.

To be consistent with Ref.~\cite{Gas05} where the meson decay constants were used
according to $F^2\rightarrow F_\pi F_K$,
we also employ $F^4 \rightarrow F_K F_\pi^3$ and 
$F^6 \rightarrow F_K F_\pi^5$ in the loop contributions, 
with $F_\pi=92.4$~MeV and $F_K=1.22\,F_\pi$.

\subsection{$\boldsymbol{K_{\ell3}}$ form factors}\label{app:Kl3formfactor}

The $K_{e3}$ form factor can be parameterized by
\beq
    f_+(t) = f_+(0)\left(1+\lambda_+\frac{t}{M_\pi^2}+\lambda_+''\frac{t^2}{M_\pi^4}
    \right) ~.
\eeq
For $f_+(0)$ we use the parameter-free one-loop result given in Ref.~\cite{Gas85I}
\beq
    f_+^{K^+\pi^0}(0) = 1.022\cdot f_+^{K^0\pi^-}(0) = 0.998 ~,
\eeq
and use the one-loop value for $\lambda_+$,
\beq
  \lambda_+ = 0.0275 ~.
\eeq
We neglect the curvature term $\propto \lambda_+''$, a final experimental conclusion
on this term seems not to have been reached so far; compare Refs.~\cite{KTeV04,Lai04}.
A similar parameterization can be used for the scalar form factor,
\beq
  f_0(t) = f_0(0)\left(1+\lambda_0\frac{t}{M_\pi^2}\right)~,
\eeq
which is related to $f_1$ and $f_+$ by 
\beq
  f_0(t) = f_+(t) + \frac{t}{M_K^2-M_\pi^2}\bigl[f_1(t)-f_+(t)\bigr] ~.
\eeq
We use 
\beq
  \lambda_0 = 0.016
\eeq
given by one-loop ChPT~\cite{Gas85I}, which is in fair agreement 
with latest experimental findings~\cite{KTeV04}.

\setcounter{equation}{0}
\section{Form factors and structure functions\\
\hspace{4.4mm}beyond the standard model}\label{app:BSMformfactors}

In this section we give expressions for the correlators
of non-standard currents defined in \eqref{eqn:BSMformfactors}
in terms of form factors and structure functions.

\subsection{Derivation of $\boldsymbol{F^s_\mu}$, $\boldsymbol{f}$, and $\boldsymbol{F^p_\mu}$}\label{app:spDeriv}

\begin{sloppypar}
Using chiral Ward identities~\cite{Bij93}, one can express the scalar correlator $F^s_\mu$ 
and the form factor $f$ in terms of $V_{\mu\nu}$ and $F_\nu$. 
This has been done in Ref.~\cite{Bra03}, so here we merely quote the results,
\beq \begin{split}
  V_{\mu\nu}W^\nu+F_\mu &= (m_u-m_s)F^s_\mu ~, \\
  F_\nu(p-p')^\nu &= (m_u-m_s)f\label{eqn:VmunuFSmu} ~.
\end{split} \eeq
Note that because of the axial anomaly, there is no corresponding relation for $F^p_\mu$. 
This pseudoscalar term can be evaluated in ChPT and contains no inner bremsstrahlung contribution. 
At leading order one obtains~\cite{Bra03,Ger04}
\begin{eqnarray}
  F^p_\mu = \frac{i\,P}{\sqrt{2}(m_u+m_s)}\,\epsilon_{\mu\lambda\rho\sigma}\,p'^\lambda q^\rho W^\sigma ~,
\end{eqnarray}
with
\begin{eqnarray}
  P = \frac{1}{2\pi^2 F^2}\frac{M_K^2}{M_K^2-W^2}\label{eqn:FpmuP} ~.
\end{eqnarray}
As pointed out in footnote~\ref{foot:scale}, 
the scalar and pseudoscalar correlators are renormalization scale dependent, 
and we have to specify (scale dependent) quark masses. 
For our numerical evaluations,
we use $m_u=4$~MeV and $m_s=115$~MeV, which are thought to be 
typical values at a scale of about 2~GeV.
\end{sloppypar}

\subsection{Decomposition of scalar and tensor correlators}\label{app:stIBSD}

The correlators $F^s_\mu$ and $\hat{T}_{\mu\nu\rho}$ 
can be split into an inner bremsstrahlung and a structure dependent part 
in a similar manner as done for $V_{\mu\nu}$, 
along the lines of Appendix~E in Ref.~\cite{Gas05}. 
For the scalar correlator
we require 
\beq
q^\mu F_\mu^{s,\IB} = f(t) ~, \qquad q^\mu F_\mu^{s,\SD} = 0 ~,
\eeq
and find
\beq \begin{split}
  F_\mu^{s,\IB} &=  \frac{p_\mu}{pq}f(W^2) + \frac{W_\mu}{qW}\Delta f ~, \\
  F_\mu^{s,\SD} &= \frac{S}{\sqrt{2}(m_u-m_s)}\bigl(qW\,p_\mu-pq\,W_\mu\bigr) ~,
\end{split} \eeq
with $\Delta f = f(t)-f(W^2)$.
Using \eqref{eqn:VmunuFSmu} we see that $S$ is given by the form factor $f_+$ and the $V_i$ as
\begin{eqnarray}
  S &=& 2 \,\frac{\Delta f_+}{qW}  + V_1 + W^2\, V_3 +p'W\, V_4 ~.
\end{eqnarray}
$F^t_{\nu\rho}$ can be parameterized as
\begin{eqnarray}
   F^t_{\nu\rho} &=& \frac{iB_T(t)}{\sqrt{2}M_K}(p_\nu p'_\rho-p'_\nu p_\rho) ~.
\end{eqnarray}
Again, we can split $\hat{T}_{\mu\nu\rho}$ into an inner bremsstrahlung part that diverges for
low photon momenta and a structure dependent part. We find
\begin{eqnarray}
    \hat{T}^{\IB}_{\mu\nu\rho} &=& \frac{i}{\sqrt{2}M_K}\bigg[
    B_T(t) \bigl(g_{\mu\nu}p'_\rho-g_{\mu\rho}p'_\nu\bigr ) \label{eqn:Thatmunurho} \\
&& + \Bigl( \frac{p_\mu}{pq}B_T(W^2) + \frac{W_\mu}{qW}\Delta B_T\Bigr) (W_\nu p'_\rho-p'_\nu W_\rho)
    \bigg] ~,  \notag
\end{eqnarray}
with $ \Delta B_T = B_T(t)-B_T(W^2)$,
and (see also Ref.~\cite{Ger04})
\begin{eqnarray}
    -i\,\hat{T}^{\SD}_{\mu\nu\rho} 
    &=& B_1\big[p'_\mu(q_\nu W_\rho-W_\nu q_\rho)-p'q(g_{\mu\nu}W_\rho-g_{\mu\rho}W_\nu)\big] \notag\\
    &+& B_2\big[p'_\mu(q_\nu p'_\rho -p'_\nu q_\rho)-p'q(g_{\mu\nu}p'_\rho-g_{\mu\rho}p'_\nu)\big] \notag\\
    &+& B_3\big(qW \,p'_\mu - p'q\,W_\mu\bigr)\bigl(p'_\nu W_\rho-W_\nu p'_\rho \bigr)\notag\\
    &+& B_4\big[W_\mu(q_\nu W_\rho - W_\nu q_\rho) \notag\\ 
    & & \qquad  - qW(g_{\mu\nu}W_\rho - g_{\mu\rho}W_\nu)\big]  \notag\\
    &+& B_5\big[W_\mu(q_\nu p'_\rho-p'_\nu q_\rho)-qW(g_{\mu\nu}p'_\rho-g_{\mu\rho}p'_\nu)\big] \notag\\
    &+& B_6\big[g_{\mu\nu}q_\rho-g_{\mu\rho}q_\nu\big] ~.
\end{eqnarray}
Little is known about the form factor $B_T$ and the structure functions $B_i$. In 
Ref.~\cite{Col99} $B_T$ is estimated to be constant and of the order of one; 
this estimate was roughly confirmed in a lattice calculation~\cite{Bec00}.
In our analysis we will use $B_T(t) = 1$. 
As we know nothing about the $B_i$ we will set them to zero.

\setcounter{equation}{0}
\section{Kinematical factors in\\\hspace{4.4mm}squared matrix elements}

In this section we quote explicit expressions for the kinematical factors that appear in the
$T$-odd parts of the squared matrix elements \eqref{eqn:Tsquared} and \eqref{eqn:TsquaredBSM},
which were obtained using \texttt{FORM}~\cite{Ver00}. 
These depend on scalar products of the external momenta that we abbreviate as follows:
\begin{xalignat}{4}
  pp' &= a~, & pq       &= b~, & pp_\ell &= c~, & pp_\nu  &= d~, \notag\\
  p'q &= e~, & p'p_\ell &= f~, & p'p_\nu &= g~, & p_\ell q &= h~, \notag\\
  p_\nu q &= j~, & p_\ell p_\nu &= k~, & pW &= l~, & p'W &= m~, \notag\\
  qW &= n~, & M_\pi^2 &= r_\pi~, & m_\ell^2 &= r_\ell~. & &
\end{xalignat}
In this appendix, all quantities are for simplicity given in units of $M_K$, i.e.\ we set $M_K=1$.
In addition, we will use $\omega = M_K^2-W^2=1-2k-r_\ell$ for
the inverse kaon propagator.
We split the normalized squared matrix elements into a $T$-even and a $T$-odd part,
\begin{eqnarray}
     \sN^{-1}\sum_{\operatorname{spins}} |T|^2 &=& |T|^2_{\even}+\xi\,|T|^2_{\odd} ~. \label{eqn:appTsquared}
\end{eqnarray}

\subsection{Hadronic loops\label{app:factorsSM}}

The $T$-odd part of \eqref{eqn:appTsquared} (see also \eqref{eqn:Tsquared}) can be written as
\begin{align}
|T|^2_{\odd} &= \notag \\
\sum_{i=1}^4 &
    \Bigl[ \Bigl(d_i f_+(t) +  e_i \delta f_+   + d_i^{II} f_1(t) + e_i^{II} \delta f_1 \Bigr) \Im{V_i} 
\notag\\
+ &\Bigl(d_i^5 f_+(t) +  e_i^5 \delta f_+   + d_i^{II5} f_1(t) + e_i^{II5} \delta f_1 \Bigr) \Im{A_i} 
\Bigr] \notag\\[2mm]
+\order&(V_i^2,A_i^2,V_iA_i) ~.\label{eqn:squaredmatrixelement}
\end{align}
We write $d_i = \hat{d}_i\bar{d}_i$ etc.\ with 
the prefactors $\hat{d}_i,\hat{e}_i,\dots$ given in Table~\ref{tab:ahat}.
Note that those multiplying $f_0(t)$, $\delta f_0$ and $\Im{V_3}$, $\Im{A_3}$ 
are proportional to $r_\ell$ and can be neglected in the electron channel.
\begin{table}
\caption{Prefactors that multiply $\bar{d}_i, \bar{e}_i$ etc.}\label{tab:ahat}    
\centering
\renewcommand{\arraystretch}{1.2}
\begin{tabular}{cccccccc}
\hline \noalign{\smallskip}
      $\hat{d}_1 \!$     & $4/(bh)$   & $\hat{d}^5_1 \!$    & $4/(bh)$            & 
                         &            & $\hat{d}^{II5}_1 \!$ & $2r_\ell/(bh)$       \\
      $\hat{d}_2 \!$     & $4/(bh)$   & $\hat{d}^5_2 \!$    & $4/(bh)$            & 
                         &            & $\hat{d}^{II5}_2 \!$ & $2r_\ell/b$          \\
      $\hat{d}_3 \!$     & $2r_\ell/h$ & $\hat{d}^5_3 \!$    & $2r_\ell/(\omega h)$ &
      $\hat{d}^{II}_3 \!$ & $r_\ell/h$  & $\hat{d}^{II5}_3 \!$ & $r_\ell/(\omega h)$  \\
      $\hat{d}_4 \!$     & $2/h$      & $\hat{d}^5_4 \!$    & $2/h$               &
      $\hat{d}^{II}_4 \!$ & $r_\ell/h$  & $\hat{d}^{II5}_4 \!$ & $r_\ell/h$           \\
\noalign{\smallskip} \hline \noalign{\smallskip}
      $\hat{e}_1 \!$     & $4/b$      & $\hat{e}^5_1 \!$    & $4/b$               &
                         &            & $\hat{e}^{II5}_1 \!$ & $2r_\ell/b$          \\
      $\hat{e}_2 \!$     & $4/b$      & $\hat{e}^5_2 \!$    & $4/b$               &
                         &            & $\hat{e}^{II5}_2 \!$ & $2r_\ell/b$          \\
\noalign{\smallskip} \hline \noalign{\smallskip}
\end{tabular}
\renewcommand{\arraystretch}{1.0}
\end{table}
The non-vanishing coefficients turn out to be
\beq \begin{array}{rclrclrcl}
\bar{d}_1 &=& \mc{7}{b(f+2h)+h(l-n-1)~,} \\[2mm]
\bar{d}_2 &=& \mc{4}{h(n-l-b)+b(k+r_\ell)~,} \quad &
\bar{d}_3 &=& \mc{1}{ e~, } \\[2mm]
\bar{d}_4 &=& \mc{7}{2g(n-b)+nr_\pi~,} \\[2mm]
\bar{d}^5_1 &=& \mc{7}{(f-g) h-b f~,} \\[2mm]
\bar{d}^5_2 &=& \mc{4}{(f-g) h+b (h+k)~,} \quad &
\bar{d}^5_3 &=& \mc{1}{e-2g~,} \\[2mm]
\bar{d}^5_4 &=& \mc{7}{-4k^2-2kr_\ell -4gk-2nk+2k-nr_\ell+n} \\[2mm]
            & & \mc{7}{-2(b+f)(g+n+2k)-2fg ~,} \\[2mm]
\bar{d}^{II}_3 &=& n~, \qquad &
\bar{d}^{II5}_1 &=& h-b~, \qquad &
\bar{d}^{II5}_3 &=& -(n+2 k)~, \\[2mm]
\bar{d}^{II}_4 &=& e~, \quad & 
\bar{d}^{II5}_2 &=& 1~, \quad &
\bar{d}^{II5}_4 &=& -(e+2g)~, \\[2mm]
\bar{e}_1 &=& \mc{7}{-b(f+g+2n)-n(l-n-1)~,}  \\[2mm]
\bar{e}_2 &=& \mc{7}{n(l-n)-b(2k+r_\ell)~, } \\[2mm]
\bar{e}^5_1 &=& \mc{4}{e(f-g)~,} &
\bar{e}^{II5}_1 &=& e~, \\[2mm]
\bar{e}^5_2 &=& \mc{4}{n(g-f)~,} &
\bar{e}^{II5}_2 &=& -n ~. 
\end{array} \eeq

\subsection{Beyond the standard model} \label{appendix:CBSM}
  Here we show the explicit expressions for the factors 
  $C_v$, $C_a$, $C_s$, $C_p$, and $C_t$ that appear in the
  squared matrix element (\ref{eqn:TsquaredBSM}).
  We do not consider the parts of $C_i$ that are quadratic in structure dependent terms.
  We have checked numerically that neglecting these terms does not change the
  final results by more than a couple of percent at most.

\paragraph{Vector and axial vector interaction.}
\newcommand{\Bv}{\alpha}
$C_v=C_a$ is decomposed according to
\beq \begin{split}
C_v = C_a = & \,\frac{1}{bh}\sum_{i=1}^4
      \Big[ \Bv_{+}^{(i)} \, f_+(t) + \tilde\Bv_{+}^{(i)} \, \delta f_+  \\
& +\,r_\ell\,\Bigl(\Bv_{1}^{(i)} \,f_1(t) + \tilde\Bv_{1}^{(i)}\,\delta f_1 \Bigr)\Big]\rea{i} ~,
\end{split} \label{eqn:Cvadecomp} \eeq
with
\beq \begin{array}{rclrcl}
    \Bv_{+}^{(1)} &=& \mc{4}{4\bigl[bf+(g-f)h\bigr]~,} \\[2mm]
    \Bv_{+}^{(2)} &=& \mc{4}{-4\bigl[(f-g)h+b(h+k)\bigr]~,} \\[2mm]
    \Bv_{+}^{(3)} &=& \mc{4}{2b(2g-e)r_\ell/\omega~,} \\[2mm]
    \Bv_{+}^{(4)} &=& \mc{4}{2b\bigl[-n-2k+4gk+2nk+4k^2+2b(g+2k+n)} \\[2mm]
	          & & \mc{4}{+2f(2g+2k+n)+(2k+n)r_\ell\bigr]~,} \\[2mm]
    \Bv_{1}^{(1)} &=& 2(b-h)~, &
    \Bv_{1}^{(2)} &=& -2h~,\\[2mm]
    \Bv_{1}^{(3)} &=& b(2k+n)/\omega~,& 
    \Bv_{1}^{(4)} &=& b(e+2g)~, \\[2mm]
    \tilde\Bv_{+}^{(1)} &=& -4eh(f-g)~, \quad &
    \tilde\Bv_{+}^{(2)} &=& 4hn(f-g)~, \\[2mm]
    \tilde\Bv_{1}^{(1)} &=& -2eh~, &
    \tilde\Bv_{1}^{(2)} &=& 2hn~. 
\end{array} \label{eqn:Cvacoeff} \eeq
In terms of powers of photon momenta $q$, $C_v=C_a$ has to be of $\order(q^{-1})$,
as it indeed proves to be according to \eqref{eqn:Cvadecomp}, \eqref{eqn:Cvacoeff}:
it arises as an IB--SD interference contribution to $\xi|T|^2_{\odd}$,
where $\xi$ itself is of $\order(q)$.

\begin{sloppypar}
\paragraph{Scalar interaction.}
\newcommand{\Bs}{\sigma}
We write $C_s$ in the form
  \begin{eqnarray}
    C_s &=& \frac{m_\ell}{m_u-m_s} \,\frac{1}{bh}\biggl\{
      \biggl[ \Bs_{+} \bigl(f_+(t)\, \delta f_0-f_0(t)\,\delta f_+\bigr)\notag\\[2mm]
        &&\qquad\qquad +~
        \Bs_{1} \bigl(f_1(t)\,\delta f_0 - f_0(t) \,\delta f_1\bigr)\notag \\[1mm]
        &&\; +~ \sum_{i=1}^4\Bigl(
        f_0(t)\,\bigl(\Bs_V^{(i)}\rev{i}+\Bs_A^{(i)}\rea{i}\bigr)\notag\\
        &&\qquad\qquad +\,\delta f_0 \,
        \tilde{\Bs}_A^{(i)}\rea{i}\Bigr)\biggr](1-r_\pi)\notag\\
      &&\; +~ \Bigl[\Bs_S^{(+)} f_+(t) + \Bs_S^{(1)} f_1(t)\Bigr] \Re{S}
    \biggr\} ~,\label{eqn:Cs}
  \end{eqnarray}
where the non-vanishing coefficients are given by
  \begin{xalignat}{2}
    \Bs_{+} &= 2e~, &
    \Bs_{1} &= n~, \notag\\
    \Bs_V^{(3)} &= bn\,~,         & \Bs_V^{(4)} &= be\,~,           \notag\\
    \Bs_A^{(1)} &= -2(b-h)~,         & \Bs_A^{(2)} &= 2h~,                \notag\\
    \Bs_A^{(3)} &= -b(2k+n)/\omega~,        & \Bs_A^{(4)} &= -b(e+2g)~,      \notag\\
    \tilde{\Bs}_A^{(1)} &= 2eh~,     & \tilde{\Bs}_A^{(2)} &= -2hn~,          \notag\\
    \Bs_S^{(+)} &= -2be~,      & \Bs_S^{(1)} &= -bn~. \label{eqn:Cscoeff}
  \end{xalignat}
We find from \eqref{eqn:Cs}, \eqref{eqn:Cscoeff} that the leading contribution to $C_s$
in the soft photon limit is again of $\order(q^{-1})$, 
although there are IB--IB interference terms.  
A stronger divergence in these is forbidden by the requirement that a $\xi$-odd
infrared singularity has to be absent,
therefore the purely bremsstrahlung contributions to the asymmetries 
\eqref{eqn:AsDecomp} are not bigger 
(or indeed numerically a bit smaller) than those proportional
to the axial structure functions.
Coefficients of $\Re{V_i}$ and $\Re{S}$ are suppressed by an additional power of $q$.
\end{sloppypar}

\paragraph{Pseudoscalar interaction.} 
\newcommand{\Bp}{\pi}
We decompose $C_p$ in the form
\beq
    C_p = \frac{m_\ell}{m_u+m_s}  \,\frac{1}{h}\Bigl[
       \Bp_{+}\,f_+(t) + \Bp_{1} \,f_1(t)\Bigr] \Re{P} ~, 
\eeq
  where
  \begin{xalignat}{2}
    \Bp_{+} &= 2(e-2g)~, & \Bp_{1} &= - (2k+n)~.
  \end{xalignat}

\paragraph{Tensor interaction.}
\newcommand{\Bt}{\theta}
We set $B_T=1$ and disregard structure dependent tensor terms. We then find
\begin{eqnarray}
     C_t &=& \frac{m_\ell}{bh} \bigg[
     \Bt_{+}\, f_+(t) + \tilde\Bt_{+}\, \delta f_+ +
     \Bt_{1}\, f_1(t) + \tilde\Bt_{1}\, \delta f_1
\notag\\ && \quad + ~
     \sum_{i=1}^4 \Bigl(\Bt_V^{(i)}\Re{V_i}+\Bt_A^{(i)}\Re{A_i}\Bigr)
     \biggr] ~, \qquad
  \end{eqnarray}
with
\beq \begin{array}{rclrclrcl}
    \Bt_{+}   &=& 4e~, \quad  & 
    \tilde\Bt_{+} &=& \mc{4}{4\bigl[b(b+m)-n(1-l)\bigr]~,} \\[2mm]
    \Bt_{1}   &=& 2n~, \quad & 
    \tilde\Bt_{1} &=& \mc{4}{2\bigl[n(e+f-g)+(n-b)r_\ell\bigr]~,} \\[2mm]
    \Bt_V^{(1)} &=& \mc{7}{4\bigl[bf+h(2b+l-n-1)\bigr]~,} \\[2mm]
    \Bt_V^{(2)} &=& \mc{7}{4\bigl[b(k+r_\ell)-h(b+l-n)\bigr]~,} \\[2mm]
    \Bt_V^{(3)} &=& \mc{7}{2b\bigl[n(g+n-f-b)+(b-n)r_\ell\bigr]~,} \\[2mm]
    \Bt_V^{(4)} &=& \mc{7}{2b\bigl[n(1-l)-b(b+m)\bigr]~,} \\[2mm]
    \Bt_A^{(1)} &=& \mc{4}{4b(b-g-n)~,} \qquad &
    \Bt_A^{(2)} &=& 4bk~,\\[2mm]
    \Bt_A^{(3)} &=& \mc{7}{2b\bigl[(n+2k)(e+f-g)+(b-n-2g)r_\ell\bigr]/\omega~,} \\[2mm]
    \Bt_A^{(4)} &=& \mc{7}{2b\bigl[b(b+f-g-4(n+k))+n(1-3f-g+n)} \\[2mm]
	        & & \mc{7}{-2mg +2k(1-2m-n-2k)-(n+2k)r_\ell\bigr]~.} 
\end{array} \vspace{2mm} \eeq 
As for the scalar interaction, both IB--IB and IB--SD interference terms
show a leading behavior $\propto 1/q$ in the soft photon limit; here, however,
the IB--IB terms turn out to be numerically dominant.

\end{appendix}


\end{document}